\documentclass[a4paper,12pt,useAMS,usenatbib]{mn2e}
\usepackage[table]{xcolor}
\usepackage{natbib}
\usepackage{sidecap}
\usepackage{float}
\usepackage{wrapfig}
\restylefloat{figure}
\usepackage{rotating}
\usepackage{amsmath}
\usepackage{longtable}
\usepackage{graphicx}
\usepackage{times}
\usepackage{psfig}
\usepackage{multicol}
\usepackage{revnum}
\usepackage[small]{caption}

\newcommand{\msun}{M$_\odot$}
\newcommand{\rsun}{R$_\odot$}
\newcommand{\kms}{km~s$^{-1}$}

\newcommand{\mytilde}{\raise.17ex\hbox{$\scriptstyle\mathtt{\sim}$}}

\title[The Binary Fraction of Planetary Nebula Central Stars I.]{The binary fraction of planetary nebula central stars\\ I. A high-precision, $I$-band excess search}
\author [O. De Marco et al.]
{\parbox{\textwidth}{Orsola De Marco$^{1,2,3}$\thanks{E-mail: \texttt{orsola.demarco@mq.edu.au}}, 
Jean-Claude Passy$^{4,3}$,
D.J. Frew$^{1,2}$,
Maxwell Moe$^5$\\
\&
G.H. Jacoby$^6$}
\vspace{0.6cm}\\
\parbox{\textwidth}{
$^1$Department of Physics \& Astronomy, Macquarie University, Sydney, NSW 2109, Australia\\
$^2$Astronomy, Astrophysics and Astrophotonics Research Centre, Macquarie University, Sydney, NSW 2109, Australia\\
$^3$Department of Astrophysics, American Museum of Natural History, New York, NY 10024, USA\\
$^4$Department of Physics and Astronomy, University of Victoria, Victoria, Canada\\
$^5$Department of Astronomy, Harvard University, Cambridge, MA 02138, USA\\
$^6$Giant Magellan Telescope and Carnegie Observatories, Pasadena, CA 91101, USA\\
}}

\begin{document}
\label{firstpage}

\maketitle

\begin{abstract}

We still do not know what causes aspherical planetary nebula morphologies. A plausible hypothesis is that they are due to the presence of a close stellar or substellar companion. So far, only $\sim$40 binary central stars of planetary nebula have been detected, almost all of them with such short periods that their binarity is revealed by photometric variability. Here we have endeavoured to discover binary central stars at any separation, thus determining the unbiased binary fraction of central stars of planetary nebula. This number, when compared to the binary fraction of the presumed parent population can give a first handle on the origin of planetary nebulae. By detecting the central stars in the $I$ band we have searched for cool companions. We have found that 30\% of our sample have an $I$ band excess detected between one and a few $\sigma$, possibly denoting companions brighter than M3-4V and with separations smaller than $\sim$1000~AU.  By accounting for the undetectable companions, we determine a de-biased binary fraction of 67-78\% for all companions at all separations. We compare this number to a main sequence binary fraction of (50$\pm$4)\% determined for spectral types F6V-G2V, appropriate if the progenitors of today's PN central star population is indeed the F6V-G2V stars. The error on our estimate cannot be constrained tightly, but we determine it to be between 10 and 30\%. We conclude that the central star binary fraction may be larger than expected from the putative parent population. However, this result is based on a sample of 27 bona fide central stars and should be considered preliminary. The success of the $I$ band method rests critically on high precision photometry and a reasonably large sample. From a similar analysis, using the more sensitive $J$ band of a subset of 11 central stars, the binary fraction is 54\% for companions brighter than $\sim$M5-6V and with separations smaller than about 900~AU. De-biassing this number in the same way as was done for the $I$ band we obtain a binary fraction of 100-107\%. The two numbers should be the same and the discrepancy is likely due to small number statistics. Finally, we note how the previously-derived short period PN binary fraction of 15-20\% is far larger than expected based on the main sequence binary fraction and period distribution. 

As a byproduct of our analysis we present an accurately vetted compilation of observed main sequence star magnitudes, colours and masses, which can serve as a reference for future studies. We also present synthetic colours of hot stars as a function of temperature (20-170kK) and gravity ($\log g= 6-8$) for Solar and PG1159 compositions.

\newpage
\end{abstract}



\newpage

\section{Introduction}
\label{sec:introduction}

A single star may be incapable of generating non-spherical planetary nebulae (PN). Models that can reproduce elliptical and bipolar PN shapes \citep[e.g.][]{GarciaSegura1999,GarciaSegura2005} have traditionally assumed the constancy of magnetic fields over the high mass-loss period (known as the superwind phase) that characterises the end of the asymptotic giant branch (AGB) evolution. However, even a weak magnetic field during the end of the AGB can act to slow down the differential rotation that generates the field in the first place: today we have no viable theory to sustain a magnetic field during the superwind phase in a single AGB star \citep{Soker2006,Nordhaus2007}. An alternative theory, that a binary companion might be responsible for the shaping action, has become central in the study of PN and formed the core of the {\it binary hypothesis} which postulates that PN form more readily around binaries, where by binary we mean a star accompanied by another star, a brown dwarf, or even a planetary system. For a review, see \citet{DeMarco2009b}.

There are several ways in which binary companions as light as planets can alter the shape of the AGB superwind and the subsequent PN \citep{Mastrodemos1999,Edgar2008,Passy2012}. However, this does not prove that all non-spherical PN (about 80\% of the entire sample; \citealt{Parker2006}) derive from a binary interaction. In order to determine the impact of binarity on PN formation a first, fundamental step is to determine the binary fraction of central stars of PN. 

Binary detection methods for central stars of PN have centred on the light variability technique, where a central, unresolved binary, undergoes eclipses, suffers ellipsoidal distortion or where the cool companion is irradiated by the luminous hot one. This technique is responsible for the detection of almost all the known central star binaries \citep{Bond2000,Miszalski2009}. The binary fraction determined in this way is $\sim$15--20\%. This technique is biased against binaries with periods smaller than about 2 weeks, against binaries with the orbital plane near the plane of the sky, and against companions with small radii \citep{DeMarco2008c}.

To determine the binary fraction for binaries with any orbital separation we need a technique that is free of separation biases, such as the detection of red and infrared excess from photometry or spectroscopy. So far only \citet{Zuckerman1991} and \citet{Frew2007} have carried out such studies -- but see also \citet{Bentley1989} and \citet{Holberg2005}. \citet{Zuckerman1991} detected definitive $K$-band excess in 50\% of 30 central stars but concluded that only in three cases this could be ascribed to a companion, while in the others the emission may be due to hot dust.  Frew \& Parker (2007) analysed 32 objects with 2-micron All Sky Survey (2MASS) or Deep Near Infrared Survey of the Southern Sky (DENIS) near-IR photometry and deduced that $>$53\% of PN have a cool companion (the completeness limit of that survey and the error limits were not quantified). 

Key to the success of such survey are (i) extremely accurate photometry with well quantified uncertainties in at least two blue colours (e.g., $B$ and $V$, to determine the reddening) and in at least one red colour ($I$ or $J$, to detect the excess flux due to the companion); (ii) a sufficiently large sample ($\ga$150 objects) covering the majority of a volume-limited sample and (iii) the use of a red band that is not contaminated by dust, practically leaving only the $I$ or $J$ bands as feasible.
Here we present the first paper in a series that uses $I$ band photometry (and $J$ where possible) to detect a red or near-IR excess. The $I$ band is not as sensitive as the $J$ band, but, if photometric errors can be limited to 1\%, this method can be a very practical way to observe a substantial number of objects in a relatively short time. 

Ideally,  we would like to derive the binary fraction and period distribution in separate samples, such as the non-spherical and the spherical subsets of PN. The need for a larger sample as well as the lack of a clear definition of morphological classes lead us to simplify this test: we simply derive the binary fraction (by which, hereafter, we intend the {\it stellar} binary fraction) among a volume limited sample of central stars of PN and compare it with that of the presumed main sequence progenitor population. The expectation from the current scenario, whereby PN derive from stars in the mass range $\sim$1-8~\msun, whether they are single or in binaries, is that the PN binary fraction should be slightly smaller than that of intermediate mass main sequence stars, if companions down to the brown dwarf limit and at all separations can be sampled. A small  difference between the binary population on the main sequence and during the PN phase is justified by the expectation that very close binary main sequence stars suffer a strong interaction on the red giant branch (RGB) and do not ascend the AGB; also, it is expected that some binary interactions on the RGB or AGB could result in mergers. 

The binary fraction of the progenitor population has recently been measured to be (50$\pm$4)\% \citep{Raghavan2010}. This includes {\it all} companions down to the planetary regime out to all separations with primary stars in the spectral range F6V to G2V. We therefore expect a PN binary fraction of approximately 10 points lower. Clearly, if we are to explain 80\% of all PN (those with non-spherical shapes) with a binary interaction, either the binary fraction of PN is higher than for the main sequence, and close to 80\%, or, if it is not, a fraction of PN should be explained by planetary interactions. While testing for the presence of planets around central stars is not generally within reach yet, we start here with determining the {\it stellar-companion}, PN binary fraction.
\begin{table*}[htbp]
\begin{center}
\scalebox{0.9}
	{\begin{tabular}{lllclclcl}
	\hline
Name   & Sp. Type & PN & D & $M_V$ & $E(B-V)$ & $T_{\rm eff}$ (method$^2$) & $\log g^3$ & Reference \& Comment \\ 
             &                & morph.$^1$ & (kpc)&(mag)&&(K)&&\\
\hline 
A~7      &  DAO  & R?&0.53    & 6.8 & 0.00   & 99$\pm$18 (m) & 7.03$\pm$0.43 & \citealt{Napiwotzki1999}; wide binary\\
A~16   &        ...   &R?                 & 2.50  & 6.3 & 0.11  & 95$\pm$8 (z He~{\sc ii}) & {\it 5.1 or 7.5}& \\  
A~20   &  O(H)    &R?            & 2.35   & 4.3 & 0.09   &119$\pm$22 (m) & 6.13$\pm$0.13 & \citealt{Rauch1999}\\ 
A~28   &   O(H)   &R?           & 1.10    & 6.2 & 0.00  & 70$\pm$30 (z He~{\sc ii})&{\it 5.1 or 7.5}&\\ 
A~31   &  DAO    &ISM        &   0.62  & 6.5 & 0.04  & 95$\pm$8 (z He~{\sc ii},m) & 6.63$\pm$0.3 & \citealt{Napiwotzki1999} \\ 
A~57   &       O(H)$^{a}$  &E/B           & 3.0: & 4.0: &0.38$^{b}$&   $>$60 (z H~{\sc i})      &  {\it 5.0}       &  $^{a}$\citealt{Miszalski2011}\\
&&&&&&&&$^{b}$\citealt{Schlafly2011}\\ 
A~71   &         ...   &ISM               & 1.22     & 7.7 &(0.8) &145$\pm$10 (z He~{\sc ii})&{\it 6.2 or 7.0}&\\
A~72   &   PG1159$^{c}$  &E       & 1.75    & 4.8 &0.09 &$>$100 (He~{\sc ii} z) & {\it 5.3 or 7.4} &  $^{c}$Douchin et al., in prep.  \\ 
A~79   &   F0 V       &T/B           &  3.30 & 2.5 & 0.62 &165$\pm$25 (Tc)&7.3& F0V central star\\
A~84   &    ...              &E            &   1.50 & 7.3 &0.16&101$\pm$10 (z He~{\sc ii}, Tc) & {\it 5.4 or 7.4} &\\
DeHt~5 & DAO            &--       &   0.35   & 7.5 &0.16&70$\pm$10 (m) & 6.65$\pm$0.19 & \citealt{Napiwotzki1999}; PN mimic\\
EGB~1  & DA     &ISM?                & 0.65     & 6.7 &0.44$^{b}$&147$\pm$25 (m)& 7.34$\pm$0.31&\citealt{Napiwotzki1999}; PN mimic?\\
EGB~6  &   DAO   &ISM            &  0.60       & 7.0 &0.03$^{b}$&110$\pm$10 (m) & 7.36 &\citealt{Gianninas2010}\\
HaWe~5  &  DA     &--           &  0.42        & 8.9 &0.20&38$\pm$2 (m)& 7.58$\pm$0.20&\citealt{Napiwotzki1999}; PN mimic\\
HDW~3   & DAO    &ISM          &0.80       &7.0 & 0.23&125$\pm$28 (m)&6.75$\pm$0.32&\citealt{Napiwotzki1999}\\
HDW~4   &   DA      &--         &    0.21     & 9.6 &0.08&47$\pm$2 (m) & 7.93$\pm$0.16 &\citealt{Napiwotzki1999}; PN mimic\\
IsWe~1    &PG1159   &ISM             &  0.63 & 7.0&...&100$\pm$20 (m)&7.0&\citealt{Werner1995}\\
IsWe~2    &  DA           &E     &   0.95    & 7.3 &...&148$\pm$20 (z H~{\sc i}) & {\it 5.4 or 7.4}&\\
JnEr~1     &    PG1159    &E/B        & 1.15 & 6.8 &0.06&130$\pm$15 (m) & 7.0 &\citet{Rauch1995}\\
K~1-13         &     ...  &E/B     &   2.45   & 6.4 &0.03& 80$\pm$30 (z He~{\sc ii}) &{\it 5.4 or 7.4}&\\
K~2-2       & hgO(H)   &ISM?        &     0.80     & 4.7 &0.03&69$\pm$15 (m) & 6.09$\pm$0.24&\citealt{Napiwotzki1999}; PN mimic?\\
NGC 3587 &hgO(H)   &E       & 0.78        & 6.3 &0.00&95$\pm$8 (m) & 6.94$\pm$0.31&\citealt{Napiwotzki1999}\\ 
NGC 6720 &DAO       &E/B          & 0.74     & 6.3 &0.08&125$\pm$25 (m)& 6.88$\pm$0.26&\citealt{Napiwotzki1999}\\ 
NGC 6853 &   DAO     &E/B               &  0.41       & 5.9 &0.07&130$\pm$10 (m) &6.72$\pm$0.23&\citealt{Napiwotzki1999}\\ 
PuWe 1      & DAO      &R/B?         & 0.37        & 7.4 &0.09&99$\pm$10 (m)&7.09$\pm$0.24&see \S~\ref{ssec:PuWe1}\\ 
Sh~2-78   & PG1159  &E/B     &   0.75      &  7.3  &0.48&120$\pm$15(m)&7.4& \citealt{Dreizler1999}\\
Sh~2-176 & DA        &ISM         &    1.0:    &  7.9  &...&150$\pm$25 (z He~{\sc ii})& {\it 6.8}&\citealt{Gianninas2010}\\
Sh~2-188 & DAO     &ISM          & 0.83       & 6.9   &0.28&130$\pm$30 (m)&6.82$\pm$0.6&\citealt{Napiwotzki1999}\\
Ton~320  &    DAO     &ISM                & 0.55        & 7.0  &0.00&78$\pm$15 (m)& 7.76 &see \S~\ref{ssec:Ton320}\\
WeDe~1  &   DA      &ISM              & 0.79         & 7.5  &    ...        &141$\pm$19 (m)&7.53$\pm$0.32& \citealt{Napiwotzki1999}\\ 
\hline
\multicolumn{9}{l}{$^1$ Legend: R: round; E: elliptical; B: bipolar; T: torus; ISM: features dominated by interaction with the interstellar medium.}\\
\multicolumn{9}{l}{$^2$ Legend: m: spectral model; z He~{\sc ii}: Zanstra method on helium lines; Z H~{\sc i}: Zanstra method on hydrogen lines; Tc: cross-over}\\ \multicolumn{9}{l}{(Ambartsumyan) temperature.}\\
\multicolumn{9}{l}{$^3$ Values in Roman font are from stellar atmosphere models, while those in italics are derived from stellar evolutionary tracks, where}\\ 
\multicolumn{9}{l}{we have chosen the higher of the two values for our calculations.}\\ 
\end{tabular}}
\caption{Observational data discussed in \S\S~\ref{sec:sample} and \ref{sec:individualobjects} \label{tab:data}}
\end{center}
\end{table*}

In \S~\ref{sec:sample} we describe our sample. In \S~\ref{sec:observations} we present our observations and data reduction, while in \S~\ref{sec:measurements} we present our measurements of the photometric magnitudes and their uncertainties. In \S~\ref{sec:technique} we give the details of the technique to detect $I$ and $J$ band excess flux and the predicted accuracies and biases that derive from it. In \S~\ref{sec:results} we report our results, including a discussion of objects that were detected to be variable in the course of our observations. A comparison between our results and the prediction for the single and binary scenarios follows in \S~\ref{sec:prediction}. In \S~\ref{sec:individualobjects} we report details of individual objects. We summarise and conclude in \S~\ref{sec:conclusions}.


\section{The sample}
\label{sec:sample}

The goal of this survey is to determine accurate $B, V$ and $I$ band photometry of the $\sim$200 central stars closest to the Sun. This sample has been recently compiled using the best available data from the literature supplemented by an improved H$\alpha$ surface brightness--radius relation which allows one to obtain distances accurate to $\sim$20\% in most cases \citep{Frew2008b}. Obtaining the $J$ band magnitude of the central stars is also desirable but much less practical for such a large sample.

The sample presented here consists of 30 central stars of PN which were selected solely based on their low PN surface brightness (radius of the PN is larger than $\sim$25~arcsec in most cases) as well as on the faint $V$ magnitudes of their central stars. The first criterion allows us to reduce the measurement error; the second insures that fainter companions can be detected because, although faint $V$ magnitudes can imply intrinsically bright, distant objects, the large PN radius tends to select for closer objects whose faint $V$ brightness indicates that the stars are intrinsically faint.
\begin{table}
\begin{center}
\scalebox{1.0}
	{\begin{tabular}{lcccl}
	\hline
Name   & $J$& $H$&$K$ &  Data  \\ 
&(mag)&(mag)&(mag)& source\\
\hline 
A~7   &     16.10$\pm$0.08     &  16.16$\pm$0.18  & $>$15.08      &  2MASS\\
A~31 &     15.95$\pm$0.01     & 15.80$\pm$0.01  &15.67$\pm$0.01  &   UKIDSS \\ 
A~72 &     16.67$\pm$0.13     &  --         & --         &  2MASS\\ 
A~79 &     15.05$\pm$0.04      &  14.64$\pm$0.06  & 14.44$\pm$0.08 &  2MASS\\
DeHt~5 &  15.57$\pm$0.07      & 15.96$\pm$0.20 &15.58$\pm$0.22   &  2MASS\\
EGB~1  &  16.64$\pm$0.16      & $>$17.47         & $>$15.62         & 2MASS\\
EGB~6  &   16.52$\pm$0.10     & 15.95$\pm$0.16  &16.10$\pm$0.26 &  2MASS\\
               &    16.43$\pm$0.20     & 16.08$\pm$0.09  &15.63$\pm$0.04 &  FL93 \\
HaWe~5  &   --           &--          & 17.78$\pm$0.18      & UKIDSS \\
HDW~3 & -- & 17.62$\pm$0.08  & 17.49$\pm$0.12  & UKIDSS\\
K~2-2       &  14.94$\pm$0.05     &14.99$\pm$0.06 &15.09$\pm$0.14 &  2MASS\\
NGC~3587& 16.71$\pm$0.13      &--        &--        &  2MASS \\ 
NGC~6720 & 16.40$\pm$0.20 &--&--&2MASS\\
NGC~6853& 14.75$\pm$0.05       &14.70         &14.61      & 2MASS \\ 
Sh~2-78&    --       &   --            & 17.89$\pm$0.15 &UKIDSS \\
Ton~320  &  16.59$\pm$0.16     &$>$15.96        &15.73$\pm$0.20& 2MASS  \\
\hline
	\end{tabular}}
	\label{tab:IRmags}
	\caption{Near IR magnitudes from the 2-micron All Sky Survey (2MASS) and the UKIRT Infrared Deep Sky Survey (UKIDSS) and \citet[][FL93]{Fulbright1993}}
\end{center}
\end{table}

In Tables \ref{tab:data} and 2
we report the best available literature data, as well as values determined here.  
We used trigonometric distances if available (e.g. Harris et al. 2007; Benedict et al. 2009), otherwise distances were taken from Frew (2008) and Frew et al. (2012).  From the distances and de-reddened $V$ magnitudes we derived the absolute magnitude, $M_V$. 
The effective temperatures are those derived via atmospheric spectrum fitting by several authors.  If a model was not available we adopted the temperature provided by Zanstra analyses of either the hydrogen or helium lines.  To do this we used the new $V$ photometry reported here in combination with new integrated H$\alpha$ fluxes taken from Frew (2008) and Frew et al. (2012).  Since these only provide a lower limit to the temperature for optically-thin PN, we have used additional information, where appropriate, to determine the most suitable temperature value.  The reddening values reported in Table~\ref{tab:data} are derived from data in the literature other than the stellar $B-V$ colour, i.e. from the nebular Balmer decrement or using the interstellar hydrogen column density.  In \S~\ref{sec:technique} we will derive the reddening using a comparison of the observed $B-V$ colour and that predicted for a single star of the appropriate temperature. At that time the nebular and stellar-derived values will be compared.
\begin{table*}
\begin{center}
\scalebox{0.75}
	{\begin{tabular}{lcclrcl}
	\hline
Name & RA & Dec & Exp Times& UT of the observation & Night & Comment \\ 
&&&$B,V,R,I$&&&\\
&&&(sec)&&\\
\hline 
A~7   &  05 03 07.52 & $-$15 36 22.8&   60, 60, 60, -- & 11:22  30/10/2007 & 1& No $I$ band\\
&&& 45, 45,  45, 90 & 08:19 03/11/2007 &5&\\
&&& 40, 40, 40, 80 & 08:35 05/11/2007&7&\\
A~16&06 43 55.46 & +61 47 24.7&120, 120, 120, 240&09:56 02/11/2007&4&\\
 &&&150, 150, 150, 300&07:50 03/11/2007&5&\\
&&&120, 120, 120, 240&10:59 04/11/2007&6&\\
A~20 &  07 22 57.74 & +01 45 32.8   &120, 120, 120, 240&10:49 02/11/2007&4&      \\  
&&&120, 120, 120, 120, 240&09:02 03/11/2007&5&\\
&&&100, 100, 100, 200&09:12 05/11/2007&7&\\
A~28 &  08 41 35.57 & +58 13 48.4   & 150, 150, 150, 300&10:15 03/11/2007&5&\\  
&&& 120, 120, 120, 240& 11:59 04/11/2007&6&\\
&&&120, 120, 120, 240&11:38 05/11/2007&7&\\
A~31 &  08 54 13.16 & +08 53 53.0   &45, 45, 45, 90  &11:24 03/11/2007&5& \\ 
&&&40, 40, 40, 80*2&10:44 05/11/2007&7&\\
A~57 &  19 17 05.73 & +25 37 33.4   & 100, 100, 100, 150&03:02 02/11/2007   &4&   \\ 
&&&120, 120, 120, 240&03:42 05/11/2007&7&\\
A~71 & 20 32 23.22  & +47 20 50.4   & 150, 150, 150, 300  &02:52 03/11/2007&5&\\
&&&200, 200, 200, 200&04:22 04/11/2007&6&\\
&&&250, 250, 250, 500&02:40 05/11/2007&7&\\
A~72 & 20 50 02.06  & +13 33 29.6    & 80,80,80,80  &03:24 01/11/2007&3&\\ 
&&& 60, 60, 60, 120& 04:59 02/11/2007&4&\\
&&&60, 60, 60, 120&04:33 05/11/2007&7&\\
A~79 & 22 26 17.27  & +54 49 38.2&100, 100, 100, 200&04:14 01/11/2007&3&\\  
&&&100, 100, 100, 200&05:35 02/11/2007&4&\\
&&&200, 200, 200, 350&06:19 04/11/2007&6&\\
A~84 & 23 47 44.02  & +51 23 56.9 &80, 80, 80, 100&04:51 01/11/2007&3&\\
&&&80, 80, 80, 160&06:02 02/11/2007&4&\\
&&&120, 120, 120, 240&05:24 05/11/2007&7&\\
DeHt~5& 22 19 33.73 & +70 56 03.1&50, 50, 50, 80&03:52 02/11/2007&4&\\
EGB~1 &01 07 07.59 &+73 33 23.2&80, 80, 80, 160&06:54 02/11/2007&4&\\
&&&80, 80, 80, 160&03:52 03/11/2007&5&\\
EGB~6 &09 52 58.99 &+13 44 34.9&60, 60, 60, 120&12:13 02/11/2007&4&\\
&&&60, 60, 60, 120&11:09 05/11/2007&7&\\
HaWe~5 &03 45 26.642 &+37 48 51.7&100, 100, 100, 200&08:40 02/11/2007&4&\\ 
&&&100, 100 100. 200&05:34 03/11/2007&5&\\
HDW~3 &03 27 15.44 & +45 24 20.5&120,120, 120, 240&07:40 30/10/2007&1& \\
&&& 160, 160, 160, 320 & 08:01 02/11/2007 &4&\\
&&& 100, 100, 100, 200 & 06:47 05/11/2007 &7&\\
HDW~4 &05 37 56.23 &+55 32 16.0&120, 120, 120, 240&09:30 02/11/2007&4& clear?\\
IsWe~1 &03 49 05.89 &+50 00 14.8 & 120, 120, 120, 240&05:29 03/11/2007&5&\\
&&& 100, 100, 100, 200&06:58 04/11/2007&6&\\
&&&100, 100, 100, 200&07:12 05/11/2007&7&\\
IsWe~2 & 	22 13 22.53 &+65 53 55.5&80, 80, 80, 160&03:48 01/11/2007&3&\\
&&&80, 80, 80, 160&04:13 02/11/2007&4&$I$ band done later at 4:53\\
&&&80, 80, 80, 160&03:25 03/11/2007&5&\\
&&&150, 150, 150, 250&05:47 04/11/2007&6&\\
&&&100, 100, 100, 200&04:59 05/11/2007&7&\\
JnEr~1 &07 57 51.63 &+53 25 17.0&120, 120, 120, 120&11:20 02/11/2007&4&\\
&&&100, 100, 100, 200&09:43 05/11/2007&7&\\
K~1-13 &08 06 46.50 &-02 52 34.8&150, 150, 150, 250&11:54 01/11/2007&3&\\
&&&120, 120, 120, 200&11:43 02/11/2007&4&\\
&&&150, 150, 150, 250&11:50 03/11/2007&5&\\
K~2-2 &06 52 23.17 &+09 57 55.7&30, 30, 30, 60&10:29 02/11/2007&4&\\
&&&30, 30, 30, 60&08:41 03/11/2007&5&\\ 
&&&30, 30, 30, 60&08:57 05/11/2007&7&\\
NGC~3587& 	11 14 47.73 & +55 01 08.5  &60, 60, 60, 120&12:35 02/11/2007&4&\\
&&&60, 60, 60, 120&12:27 04/11/2007&5&\\
&&&60, 60, 60, 120&12:41 05/11/2007&6&\\
NGC~6720& 	18 53 35.08 & +33 01 45.0 &50, 50, 50, 50&01:59 02/11/2007&4&\\
&&&30, 30, 30, 60&03:23 05/11/2007&7&\\
NGC~6853& 	19 59 36.38 &+22 43 15.7 &15, 15, 15, 15&03:32 02/11/2007&4&\\
&&&30,30,30,60&02:34 03/11/2007&5&\\
&&&20,20,20,40&04:13 05/11/2007&7&\\
PuWe~1& 06 19 34.33 &+55 36 42.3 &50, 50, 50, 100 & 07:25 03/11/2007&5&\\
&&&50, 50, 50, 100&07:40 05/11/2007&7&\\
Sh~2-78 &19 03 10.09 &+14 06 58.9&100, 100, 100, 150&02:32 02/11/2007&4& \\
&&&150, 120, 120, 400&01:58 03/11/2007&5&\\
Sh~2-176 &00 31 54 &+57 22.6&80, 80, 80, 160&05:17 01/11/2007&3&\\
&&&80, 80, 80, 160&06:27 02/11/2007&4&\\
&&&150, 150, 150, 300&05:50 05/11/2007&7&\\
Sh~2-188 &01 30 33.11 &+58 24 50.7&120, 120, 120, 240&07:16 02/11/2007&4&\\
&&&120, 120, 120, 240&04:17 03/11/2007&5&\\
&&&120, 120, 120, 240&06:21 05/11/2007&7&\\
Ton~320 &08 27 05.53 &+31 30 08.6&60, 60, 60, 120&09:42 03/11/2007&5&\\
&&&60, 60, 60, 180&10:15 05/11/2007&7&\\
WeDe~1 &05 59 24.87 &+10 41 40.4&120, 120, 120, 240&06:53 03/11/2007&5\\
&&&120, 120, 120, 240 & 11:31 04/11/2007&6&\\
&&&120, 120, 120, 240&08:05 05/11/2007&7&\\
\hline
	\end{tabular}}
	\caption{Log of observations taken in photometric conditions \label{tab:logs}}
\end{center}
\end{table*}

Out of the 30 objects studied in the $I$ band, three were later discovered to be PN mimics, nebulae which we exclude to have been ejected by the central star when it was on the AGB \citep{Frew2010c}. These are HaWe5 and HDW4, both of which have a very high surface gravity indicating a long cooling age. Such stars departed from the AGB well over the maximum lifetime of a PN of $\sim$100,000 years. Their nebulae must be Str\"omgren spheres ionised by the hot central stars. In addition we also consider DeHt~5 a PN mimic due to a range of discrepancies which we describe fully in \S~\ref{ssec:DeHt5}. Although we suspect that two other objects are also PN mimics (EGB~1 and and K~2-2), for the moment we keep them in the central star sample. Our {\it bona fide} sample comprises 27 objects studied in the $I$ band of which 11 have additional $J$-band photometry. A~79 has a cool spectral type \citep{Rodriguez2001}, indicating that it is a binary, but this was not realised at the time of the observations.

In Column 3 of Table~\ref{tab:data} we have listed the morphologies of the PN. Many of them are too old and dispersed for a morphology to be determined accurately, and the PN features are dominated by interaction with the ISM. Those PN we list as ``R?" are approximately round, but may contain internal structure which, in some cases, is distinctly bipolar. An example is A~7, which, although diffuse, contains a hint of bipolarity in the form of brighter structure on either side of the star. None of the PN we mark as ``R?" is like the round PN Abell~39 \citep{Jacoby2001}. A~20 is a faint round PN with an evacuated cavity around the central star. Its appearance is that of a pole-on doughnut. A~28 on the other hand, has a sharp, broken rim and may be more elliptical than round. We will not mention morphology further. It is not good form, when trying to determine if binarity causes certain morphologies, to use the morphology as an argument for binarity.



\section{Observations and Data Reduction}
\label{sec:observations}

The observations were acquired during 8 nights between the 30th of October and the 6th of November, 2007 at the 2.1-m telescope at Kitt Peak National Observatory. However, the data from nights 2 and 8 were not photometric. The weather conditions during the other nights were mostly photometric. Data taken in non-photometric conditions were used to monitor stars for relative variability (see \S~\ref{ssec:variability}). A log of the observations is presented in Table~\ref{tab:logs}.

The detector was a $2048 \times 2048$ pixel Tetronix CCD, which was binned $2 \times 2$ (to achieve a faster read-out and reduce the read noise level). The pixel size was $24 \ \mu m$, and the field of view was $10.2' \times 10.2'$ (the platescale is 0.60 arcsec per (binned) pixel). The electronic gain of the camera was $ 3.1 \ e^{-} \rm{/ ADU}$ (Analog-to-Digital Unit), which minimally samples the system noise of 6 $e^{-}$, while providing a maximum signal of about 200,000 $e^{-}$ before saturating.

The observations were made through the $B$,$V$,$R$ and $I$ Johnson-Cousins astronomical bandpasses. 
Ten bias frames per nights were obtained as well as ten dome flats per filter. Their medians were used to de-bias and flat field the exposures.
Standard stars were selected from the list of \citet{Landolt1992}, so as to encompass the colours and brightnesses of our targets as well as to sample a range of air masses. They are listed in Table~4 alongside with the nights  when they were used. We applied a shutter correction of $-0.001$~sec to the header exposure times, very low for even our shortest exposure times.


\section{The determination of the photometric magnitudes}
\label{sec:measurements}

Instrumental magnitudes were measured with the {\sc apphot} package in {\sc iraf}\footnote{{\sc iraf} is distributed by the National Optical Astronomy Observatories, which are operated by the Association of Universities for Research in Astronomy, Inc., under cooperative agreement with the National Science Foundation.} \citep{Tody1986,Tody1993}. The radius of the aperture that samples the flux was estimated to be 8 binned pixels by measuring bright stars through increasingly larger apertures. The background was sampled between 8 and 13 binned pixels from the position of the star. Targets were measured using apertures of 3 binned pixels and then aperture-corrected. However, it was noticed that the final result was not very different from that obtained by measuring targets through an 8-binned pixel aperture directly with no aperture correction.

The instrumental magnitudes $b,v,r \ {\rm and} \ i$, were then converted to observed magnitudes, $B,V,R \ {\rm and} \ I$ on the standard system, by making the following first order transformation:

\begin{equation}
\begin{array}{lll}
	B = O_B + b + C_B(B-V) - K_B*Z_B \\
	V = O_V + v + C_V(B-V) - K_V*Z_V \\
	R = O_R + r + C_R(V-R) - K_R*Z_R \\ 
	I = O_I + i + C_I(R-I) - K_I*Z_I
\end{array}
\label{eq:instrument}
\end{equation}

\noindent where $O_B,O_V,O_R,O_I$ are the instrumental offsets,  $C_B,C_V,C_R,C_i$ are the colour terms, $K_B,K_V,K_R,K_I$ are the extinction coefficients, and $Z$ is the airmass (where the $Z$ values are all the same and the subscripts are there for clarity). There can be a second order correction due to the atmosphere, and therefore zenith distance because the atmosphere acts as a broadband colour filter. This second order correction was negligible. Although the airmass of our observations varied, it was never higher than 2.1 and generally very close to unity.

\begin{table*}
\begin{center}
\scalebox{0.8}
	{\begin{tabular}{lccccccc}
	\hline
	Coeff. & Night 1  & Night 3 & Night 4 & Night 5 & Night 6 & Night 7 \\
	\hline
	$O_{B}$ & $-0.8638  \pm 0.0083 $  & $-0.8209 \pm 0.0157$ & $-0.8436 \pm 0.0080 $  & $-0.8525 \pm 0.0088 $  & $-0.9012 \pm 0.0153$  & $-0.8466 \pm 0.0079$    \\
	$O_{V}$ &  $-0.8054 \pm 0.0073 $   & $-0.7767 \pm0.0168$  & $-0.7877 \pm 0.0100 $  & $-0.7958 \pm 0.0069 $  & $-0.8155 \pm 0.0167$  & $-0.7970 \pm 0.0064$          \\
	$O_{R}$ &  $-0.7940 \pm 0.0068 $   & $-0.6169 \pm 0.0244$  & $-0.7656 \pm 0.0127 $  & $-0.7824 \pm 0.0140 $  & $-0.7775 \pm 0.0193$  & $-0.7887 \pm 0.0127$         \\
	$O_{I}  $ &  $-1.774 \pm 0.008 $   & $-1.790 \pm 0.038$  & $-1.756 \pm 0.013 $  & $-1.778 \pm 0.017 $  & $-1.758 \pm 0.035$   & $-1.752 \pm 0.013$         \\
	\hline
	$K_{B}$ & $0.2286  \pm 0.0059  $  & $0.2550 \pm 0.0107$ & $0.2370 \pm 0.0054 $ & $0.2355 \pm 0.0056 $ & $0.2205 \pm 0.0106$ & $0.2522 \pm 0.0055$       \\
	$K_{V}$ & $0.1313  \pm 0.0050  $   & $0.1445 \pm 0.0113$ & $0.1373 \pm 0.0070 $ & $0.1325 \pm 0.0043 $ & $0.1441 \pm 0.0101$ & $0.1464 \pm 0.0042$           \\
	$K_{R}$ & $0.09011  \pm 0.00468 $   & $0.2310 \pm 0.0165$ & $0.1011 \pm 0.0091 $ & $0.08112 \pm 0.00886 $ & $0.1072 \pm 0.0178$  & $0.09376 \pm 0.00863$          \\
	$K_{I}$   & $0.04785  \pm 0.00537 $   & $0.05281 \pm 0.02654$ & $0.04264 \pm 0.00956 $ & $0.01315 \pm 0.01101 $ & $0.04623 \pm 0.02283$  & $0.04895 \pm 0.00935$          \\	
	\hline
	$C_{B}$ & $0.02832 \pm 0.00320 $ & $0.03006 \pm 0.00546$& $0.02772 \pm 0.00308 $ & $0.03248 \pm 0.00298 $ & $0.02849 \pm 0.00600$  & $0.02747 \pm 0.00273$          \\
	$C_{V}$ & $-0.03070 \pm 0.00264 $ & $-0.02954 \pm 0.00607$& $-0.03141 \pm 0.00335 $ & $-0.03480 \pm 0.00232 $ & $-0.03465 \pm 0.00670$  & $ -0.03137 \pm 0.00254$          \\
	$C_{R}$ & $-0.03581 \pm 0.00472 $ & $-0.05941\pm 0.01434$& $-0.04464 \pm 0.00754 $ & $-0.06885 \pm 0.00696 $ & $-0.1066 \pm 0.0125$ & $ -0.06955 \pm 0.00804$           \\
	$C_{I}  $ & $0.05139 \pm 0.00586 $ & $-0.2063 \pm 0.0161$& $0.03163 \pm 0.00944 $ & $-0.06926 \pm 0.01115 $ & $-0.1347  \pm 0.0258$ & $-0.02935 \pm 0.00943$           \\	
	\hline
	\# Stars  & 27,28,24,23 & 21,18,11,9 & 54, 47, 35, 27 & 46, 49, 33, 35 & 24,28,22,25 & 50,63,54,50 \\
	$(B,V,R,I)$ &&&&&&\\
	\hline
	\end{tabular}}
	\label{tab:coeff}
	\caption{Coefficients and errors for every photometric night. As explained in \S~\ref{sec:observations}, nights 2 and 8 were non-photometric}
\end{center}
\end{table*}

\begin{figure*}
\vspace{9cm}
	\begin{center}$
	      \begin{array}{cc}
		\includegraphics{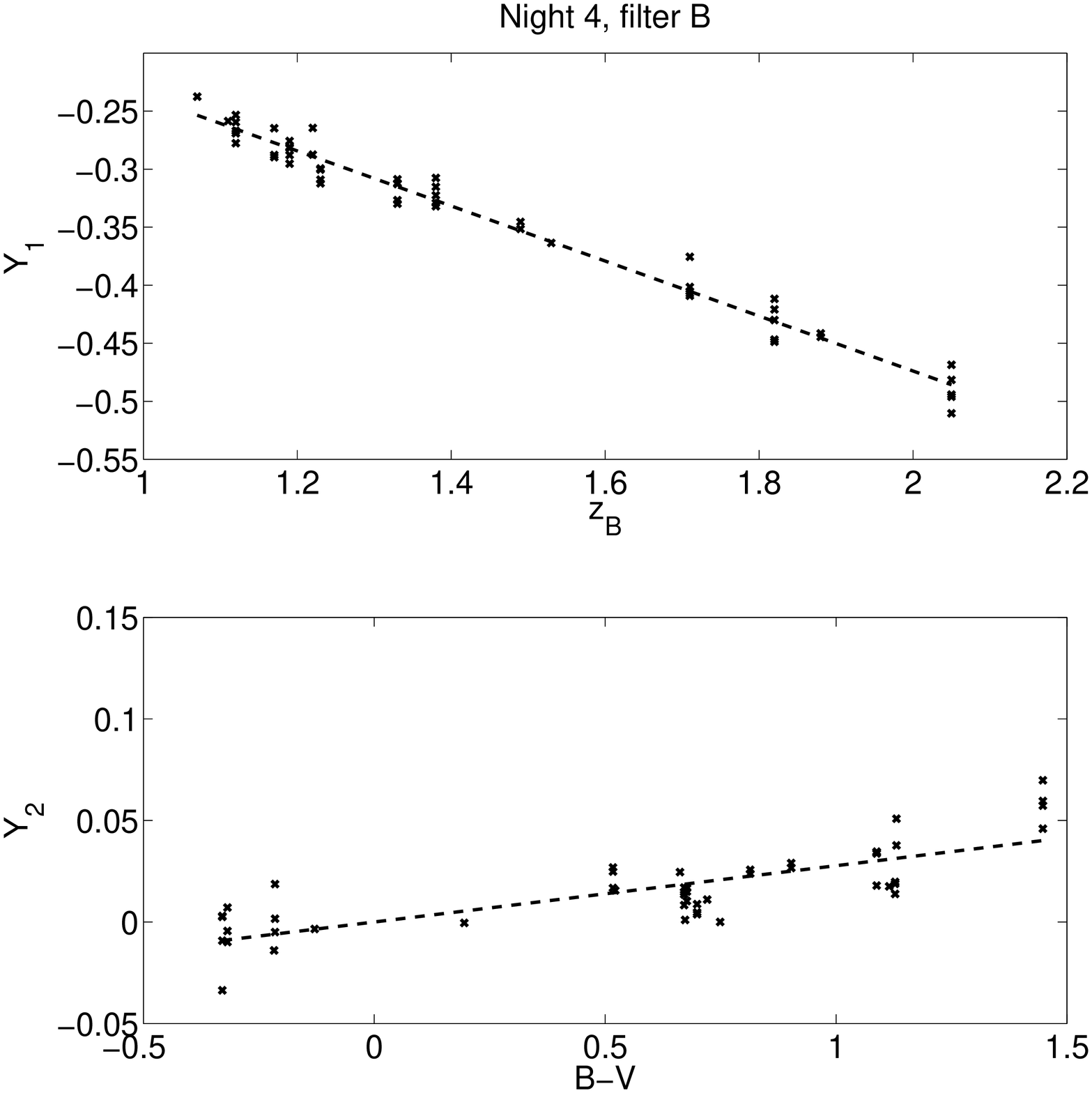} &
		\includegraphics{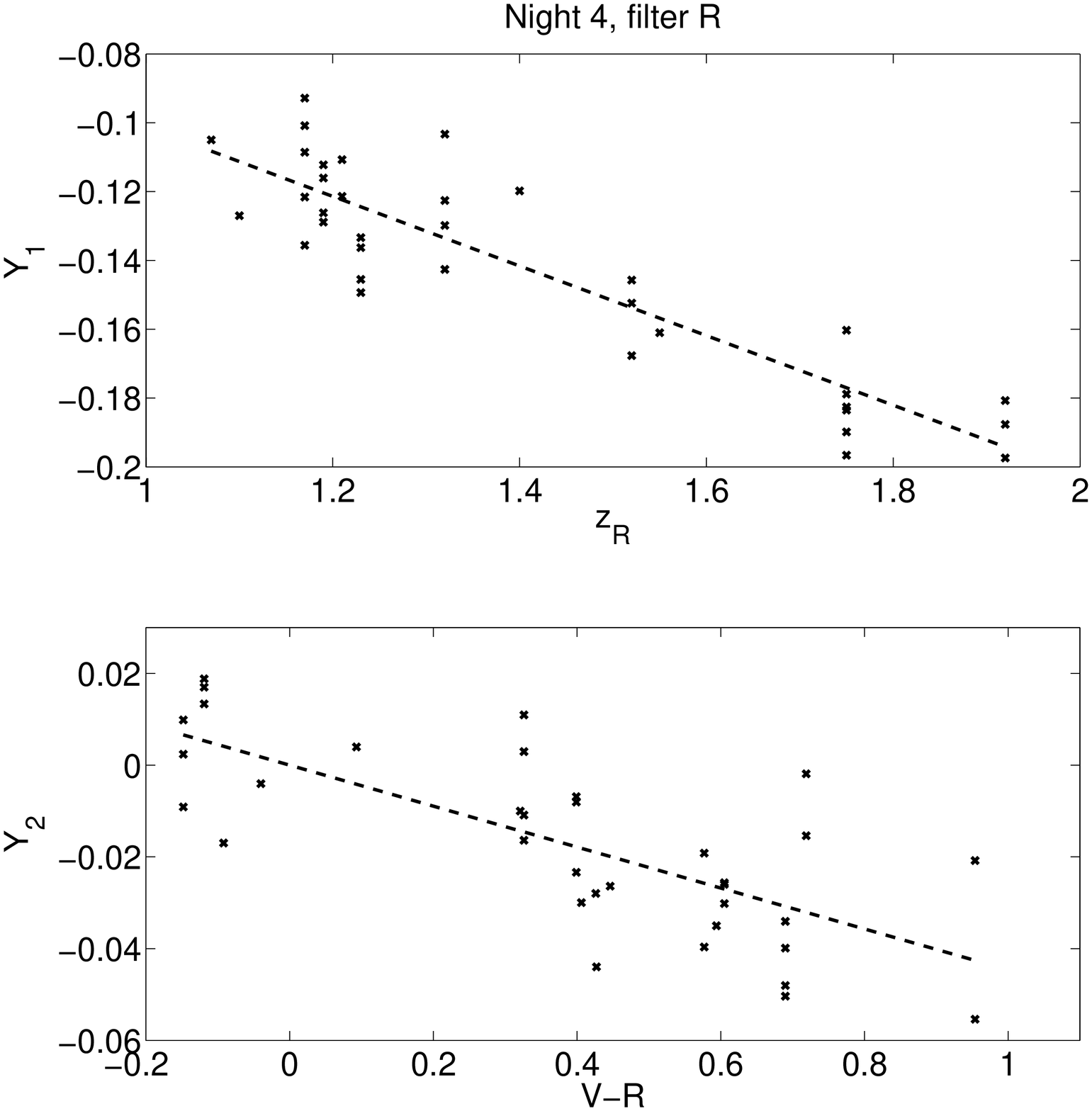}
		\end{array}$
		\caption{Example of fits to standard stars to estimate conversion coefficients. Note the small range on the $y$-axis of the $R$ band fit}
		\label{fig:standardfits}
	\end{center}
\end{figure*}

Our standard stars were used to calculate the needed coefficients.  Each standard star was observed through each filter at several values of the airmass, $Z$, and its instrumental magnitudes $b, v, r$ and $i$ were determined. Then, for each filter, we solved the system of equations in (\ref{eq:instrument}) using a least squares method, where the unknowns are $K,C \ \rm {and} \  O$. We finally plot $Y_1 =  V - O_V - v - C_V(B-V)$ as a function of $Z$, and $Y_2 = I - O_I - i + K_I*Z$ as a function of $R-I$. In Figure \ref{fig:standardfits}, we show the importance of obtaining observations of several standards at different airmass values. When there are few stars (less than $\sim$25 stars), the fits tend to be noticeably poorer. The values obtained for each coefficient as well as the total number of stars in each filter are summarised in Table~5.

\subsection{Uncertainties}
\label{ssec:uncertainties}

Several precautions are taken to minimise the uncertainty. First, the targets' signal-to-noise ratio is $\ga$100. Second, multiple observations of each target are taken on different nights. This guards against possible mistakes in estimating the photometric conditions. It also allows us to assess variability affecting 15-20\% of all central stars of PN and usually denoting binarity in its own right (see \S~\ref{ssec:variability}). Finally, most of our targets have faint, large PN ($>$25~arcsec), which reduces the error due to background subtraction. 

The uncertainties on the instrumental magnitude measurements ($\sigma_b$, $\sigma_v$, $\sigma_r$, and $\sigma_i$) provided by {\sc iraf} include the photon statistics and the uncertainty on the sky measurement. While the shutter correction was included, we did not include a shutter correction error because it was deemed negligible.
By adding the uncertainties in quadrature we have, e.g., in the case of  the error on the $B$-band magnitude:
\begin{table}
\begin{center}
	{\begin{tabular}{lcc}
	\hline
	Name  & \# stars & Night observed \\ \hline
	GD 246          &1B& 1,3,4,5,6,7 \\
	PG0231+051 & 1B + 6    & 1,3,4,5,6,7 \\
	PG0918+029 & 1B + 4   & 5 \\
	PG1047+003 & 1B + 3    & 5 \\
	PG2213-006 & 1B + 3     & 3,4,5,6,7 \\
	RU 149      &2B + 6& 1,4,5,6,7 \\
	92 245${^1}$ &4& 1,3,4,5,6,7\\
	95 112${^2}$ &1B + 4 & 1,3,4,5,6,7 \\
	98 650${^3}$ &1B + 9 & 1,3,4,5,6,7 \\
	\hline
	\multicolumn{3}{l}{${^1}$This field also contains 92~248, 92~249 and 92~250.}\\
	\multicolumn{3}{l}{${^2}$This field also contains 95~41-43 and 95~115.}\\
	\multicolumn{3}{l}{${^3}$This field also contains 98~653, 98~670-671,}\\ 
	\multicolumn{3}{l}{98~675-676, 98~L5, 98~682, 98~685 and 98~1087.}
	\end{tabular}}
	\label{tab:stdstars}
	\caption{Standard stars from \citet{Landolt1992} and \citet{Landolt2009}. In Column 2 we indicate the number of blue stars (marked ``B"), as well as of redder stars}
\end{center}
\end{table}

\begin{multline*}
 \sigma_B^2 = \\
  \sigma_{O_B}^2 + \sigma_b^2 + (B-V)^2\sigma_{C_B}^2 + C_B^2\sigma_{B-V}^2 + Z_B^2 \sigma_{K_B}^2 + \sigma_{apco}^2
\end{multline*}

\noindent where $\sigma_{apco} \approx 0.004$~mag is the error on the aperture correction, calculated as the standard deviation of the aperture correction for the 5 reference stars used to calculate the aperture correction itself. 
By adopting the approximation $\sigma_{B-V}^2 = \sigma_{B}^2 + \sigma_{V}^2 \approx 2 \sigma_{V}^2 $, we obtain the following equations for the errors on the $B$ and $V$ bands:

\begin{multline}
\begin{array}{lll}
	\sigma_B^2 = \\
	\left( \sigma_{O_B}^2 + \sigma_b^2 + (B-V)^2\sigma_{C_B}^2 + Z_B^2 \sigma_{K_B}^2 + \sigma_{apco}^2 \right) / \\
	( 1 - 2C_B^2)\\
	\\
	\sigma_V^2 = \\
	\left(\sigma_{O_V}^2 + \sigma_v^2 + (B-V)^2\sigma_{C_V}^2 + Z_V^2 \sigma_{K_V}^2 + \sigma_{apco}^2 \right) / \\
	( 1 - 2C_V^2) 
\end{array}
\label{eq:errors1}
\end{multline}

\noindent which we solve simultaneously to derive the values of $\sigma_B$ and $\sigma_V$. Similarly, we derive the uncertainties on $R$ and $I$. 

When more than one independent observation was taken for a given target, the weighted mean was calculated: 

\begin{equation}
\mu = \Sigma_{i=1}^N p_i x_i , 
\label{eq:mean}
\end{equation}

\noindent where $N$ is the number of observations, $x_i$ is a given measurement and $p_i$ is its associated probability ($p_i = (1/\sigma_i) / \Sigma_j (1/ \sigma_j)$, where $\sigma_i$ is the error on that measurement and $\sigma_j$ is the error on the j$^{\rm th}$ measurement). The error on the mean was calculated in the following way:

\begin{equation}
\sigma = \sqrt{\Sigma_{i=1}^N p_i ( x_i - \mu)^2} .
\label{eq:meanerror}
\end{equation}

\noindent The photometric magnitudes and their errors thus obtained are reported in Table~\ref{tab:results}, along with the number of measurements that were used in obtaining these values. In Appendix A we report the individual values and their individual errors.


\section{Binary detection technique by red and IR excess flux}
\label{sec:technique}

The ideal spectral location for this technique is the $J$ band. This is the best compromise between companion brightness and the elimination of  contamination from hot dust. However, procuring $J$ band data has proven more challenging than optical data because IR instruments are often associated with larger telescopes which tend to allocate shorter observing runs and because of the need for photometric conditions. In addition, for most of our sample we also need to obtain $B$- and $V$-band photometry because the values in the literature tend not to be sufficiently accurate, thus generating the need for parallel proposals to more than one telescope. As a result, we have found it more practical to use the $I$ band to detect companion-generated flux excess as our work horse, accompanied by the $J$ band when available (we used UKIRT, Infrared Deep Sky Survey (UKIDSS; \citealt{Lawrence2007}) data, or the most reliable 2MASS values -- see Table~2).
Below we discuss the technique and its biases.

\begin{table*}
\begin{tabular}{lcccc}
\hline
Name & B & V & R & I   \\ \hline \hline
A~7           & 15.190     $\pm$  0.003  (   3   )& 15.495     $\pm$  0.004  (   3   )& 15.632     $\pm$  0.013  (   3   )& 15.818     $\pm$  0.011  (   2  )\\
A~16         &18.517     $\pm$  0.015  (   3   )& 18.714     $\pm$  0.018  (   3   )& 18.733     $\pm$  0.020  (   3   )& 18.686     $\pm$  0.016  (   3  )\\
A~20        &16.216     $\pm$  0.002  (   3   )& 16.466     $\pm$  0.006  (   3   )& 16.559     $\pm$  0.007  (   3   )& 16.701     $\pm$  0.016  (   3  )\\
A~28       &16.280     $\pm$  0.008  (   3   )& 16.557     $\pm$  0.009  (   3   )& 16.691     $\pm$  0.008  (   3   )& 16.877     $\pm$  0.014  (   3  )\\
A~31        &15.201     $\pm$  0.007  (   2   )& 15.544     $\pm$  0.001  (   2   )& 15.693     $\pm$  0.009  (   2   )& 15.831     $\pm$  0.016  (   2  )\\
A~57       &17.903     $\pm$  0.001  (   2   )& 17.734     $\pm$  0.011  (   2   )& 17.451     $\pm$  0.010  (   2   )& 17.210     $\pm$  0.002  (   2  )\\
A~71        &19.384     $\pm$  0.015  (   3   )& 19.335     $\pm$  0.006  (   3   )& 19.253     $\pm$  0.010  (   3   )& 19.185     $\pm$  0.056  (   3  )\\
A~72        &15.761     $\pm$  0.021  (   3   )& 16.070     $\pm$  0.028  (   3   )& 16.237     $\pm$  0.091  (   3   )& 16.381     $\pm$  0.043  (   3  )\\ 
A~79	&17.825     $\pm$  0.013  (   3   )& 16.965     $\pm$  0.005  (   3   )& 16.397     $\pm$  0.026  (   3   )& 15.743     $\pm$  0.085  (   3  )\\
A~84		 &18.366     $\pm$  0.012  (   3   )& 18.584     $\pm$  0.013  (   3   )& 18.613     $\pm$  0.020  (   3   )& 18.671     $\pm$  0.020  (   3  )\\
DeHt~5	 &15.268     $\pm$  0.018  (   1   )& 15.495     $\pm$  0.019  (   1   )& 15.568     $\pm$  0.018  (   1   )& 15.631     $\pm$  0.018  (   1  )\\
EGB~1	 &16.308     $\pm$  0.007  (   2   )& 16.439     $\pm$  0.005  (   2   )& 16.452     $\pm$  0.007  (   2   )& 16.482     $\pm$  0.022  (   2  )\\
EGB~6	 &15.692     $\pm$  0.002  (   2   )& 15.999     $\pm$  0.002  (   2   )& 16.137     $\pm$  0.008  (   2   )& 16.300     $\pm$  0.009  (   2  )\\
HaWe~5	 &17.321     $\pm$  0.014  (   2   )& 17.439     $\pm$  0.005  (   2   )& 17.471     $\pm$  0.004  (   2   )& 17.528     $\pm$  0.011  (   2  )\\
HDW~3	  &17.084     $\pm$  0.003  (   3   )& 17.187     $\pm$  0.004  (   3   )& 17.218     $\pm$  0.039  (   3   )& 17.234     $\pm$  0.019  (   3  )\\
HDW~4	 &16.310     $\pm$  0.011  (   1   )& 16.540     $\pm$  0.013  (   1   )& 16.638     $\pm$  0.017  (   1   )& 16.739     $\pm$  0.017  (   1  )\\
IsWe~1	 &16.374     $\pm$  0.017  (   3   )& 16.523     $\pm$  0.007  (   3   )& 16.576     $\pm$  0.016  (   3   )& 16.644     $\pm$  0.013  (   3  )\\
IsWe~2	 &18.142     $\pm$  0.026  (   5   )& 18.160     $\pm$  0.033  (   5   )& 18.118     $\pm$  0.026  (   5   )& 18.098     $\pm$  0.022  (   5  )\\
JnEr~1	 &16.775     $\pm$  0.005  (   2   )& 17.128     $\pm$  0.013  (   2   )& 17.288     $\pm$  0.001  (   2   )& 17.501     $\pm$  0.023  (   2  )\\
K~1-13	 &18.051     $\pm$  0.013  (   3   )& 18.425     $\pm$  0.006  (   3   )& 18.592     $\pm$  0.019  (   3   )& 18.846     $\pm$  0.044  (   3  )\\
K~2-2	 &13.977     $\pm$  0.007  (   3   )& 14.263     $\pm$  0.010  (   3   )& 14.390     $\pm$  0.008  (   3   )& 14.553     $\pm$  0.013  (   3  )\\
NGC~3587	 &15.414     $\pm$  0.001  (   3   )& 15.777     $\pm$  0.009  (   3   )& 15.960     $\pm$  0.006  (   3   )& 16.194     $\pm$  0.029  (   3  )\\
NGC~6720	 &15.405     $\pm$  0.016  (   2   )& 15.769     $\pm$  0.023  (   2   )& 15.901     $\pm$  0.003  (   2   )& 16.062     $\pm$  0.012  (   2  )\\
NGC~6853	 &13.749     $\pm$  0.026  (   3   )& 14.089     $\pm$  0.010  (   3   )& 14.247     $\pm$  0.006  (   3   )& 14.405     $\pm$  0.010  (   3  )\\
PuWe~1	 &15.291     $\pm$  0.008  (   2   )& 15.545     $\pm$  0.006  (   2   )& 15.662     $\pm$  0.011  (   2   )& 15.792     $\pm$  0.008  (   2  )\\
Sh~2-78	 &17.633     $\pm$  0.012  (   2   )& 17.660     $\pm$  0.005  (   2   )& 17.608     $\pm$  0.025  (   2   )& 17.543     $\pm$  0.027  (   2  )\\
Sh~2-176		 &18.489     $\pm$  0.086  (   3   )& 18.559     $\pm$  0.019  (   3   )& 18.570     $\pm$  0.016  (   3   )& 18.545     $\pm$  0.029  (   3  )\\
Sh~2-176$^1$ &18.442     $\pm$  0.018  (   2   )& 18.551     $\pm$  0.016  (   2   )& 18.562     $\pm$  0.009  (   2   )& 18.531     $\pm$  0.010  (   2  )\\
Sh~2-188		 &17.424     $\pm$  0.013  (   3   )& 17.447     $\pm$  0.004  (   3   )& 17.398     $\pm$  0.009  (   3   )& 17.376     $\pm$  0.011  (   3  )\\
Ton~320	 &15.379     $\pm$  0.007  (   2   )& 15.725     $\pm$  0.006  (   2   )& 15.890     $\pm$  0.010  (   2   )& 16.105     $\pm$  0.018  (   2  )\\
WeDe~1	 &16.958     $\pm$  0.007  (   3   )& 17.226     $\pm$  0.004  (   3   )& 17.338     $\pm$  0.010  (   3   )& 17.489     $\pm$  0.016  (   3  )\\
\hline
\multicolumn{5}{l}{$^1$These measurements are obtained by excluding the observations taken in night 3.}
\end{tabular}
\caption{The photometric magnitudes of our targets rounded to 3 decimal places.  Formal errors lower than 1\% were set at 1\%. In brackets are the number of independent exposures (taken on different nights) used to calculate the final photometric magnitude and the uncertainty. In Table~\ref{tab:individualmagnitudes} we present the individual magnitudes whose weighted averages are displayed here \label{tab:results}}
\end{table*}
Most of our targets have reasonably well determined effective temperatures either via stellar spectrum modelling or using the Zanstra technique (Table~\ref{tab:data}). Those that were modelled also have known gravities. For those stars with only a Zanstra temperature estimate, we determine the gravity using the most common central star mass, namely 0.61-\msun, selecting the corresponding stellar evolutionary track from fig.~2 of \citet[][who used tracks from \citet{Schoenberner1983}, \citet{Koester1986} and \citet{Bloecker1995}]{Napiwotzki1999} and choosing the larger of the possible gravity values, appropriate for our sample of evolved PN. Once the temperatures are obtained, single star colours  ($B-V$, $V-I$ and $V-J$) can be predicted. The predicted $B-V$ colour is used together with the measured one to determine the reddening, $E(B-V)$. If the reddening thus determined has a negative value, possible because of random errors, we set the value to zero. As we will see in \S~\ref{sec:results}, all negative reddening values we derive in this way are very small and within the uncertainty, reassuring us that the uncertainties have been assessed reasonably. With the reddenings thus derived we obtain values of $A_\lambda/E(B-V)$ using the reddening law of \citet[][]{Cardelli1989}, where $\lambda$ represent the bandpass central wavelengths. The central wavelengths for the filter bandpasses are obtained by convolving the bandpasses with a synthetic stellar atmosphere with $T_{\rm eff}$=100~kK, $\log g=7$ and solar abundance. This decreases the central wavelengths by approximately 10~\AA, compared to the values obtained for un-convolved bandpasses, but yields only a very small change in the final results. The bandpass central wavelengths thus determined and the values of the wavelength-specific extinctions are presented in Table~\ref{tab:extinctions}.

The $V-I$, or $V-J$ observed dereddened colours are then compared with the predicted ones so as to determine if a flux excess exists. An $I$ ($J$) band excess is detected every time the observed (dereddened) and predicted colours are different by more than the combined uncertainties. While a high confidence result can already lend substantial weight towards a binary interpretation, a lower sigma result needs to be confirmed by additional photometry or spectroscopy.

To predict the $B-V$, $V-I$, $V-J$, $R-I$ and $J-H$ colours of single post-AGB stars, we use theoretical stellar atmosphere models calculated with the simulation code TMAW, the web interface to TMAP \citep{Werner1999,Werner2003,Rauch2003}, or the German Astrophysical Virtual Observatory grid calculations TheoSSA\footnote{dc.zah.uni-heidelberg.de/theossa/}. The colours and further details of the calculations are provided in Appendix~\ref{app:stellarmodels}. 

The uncertainties on the measurements (see \S~\ref{ssec:uncertainties}) are combined with the uncertainty on the predicted colours to derive an uncertainty on the colour excess. The uncertainty on the theoretical colours reflect solely the uncertainty on the temperature. We assume that additional sources of uncertainty are far smaller and do not play a role. The uncertainty in colours derived from the uncertainty on the temperature is quite small, in particular for stars hotter than $\sim$50~kK, which is the case for most of our sample. Systematic uncertainties on the theoretical, single star colours are estimated to be below 1\% \citep[e.g.,][]{Rauch2007}. In fact, the observational uncertainties dominate the error budget. 

\begin{figure*}
\vspace{6cm}
	\begin{center}
		\includegraphics{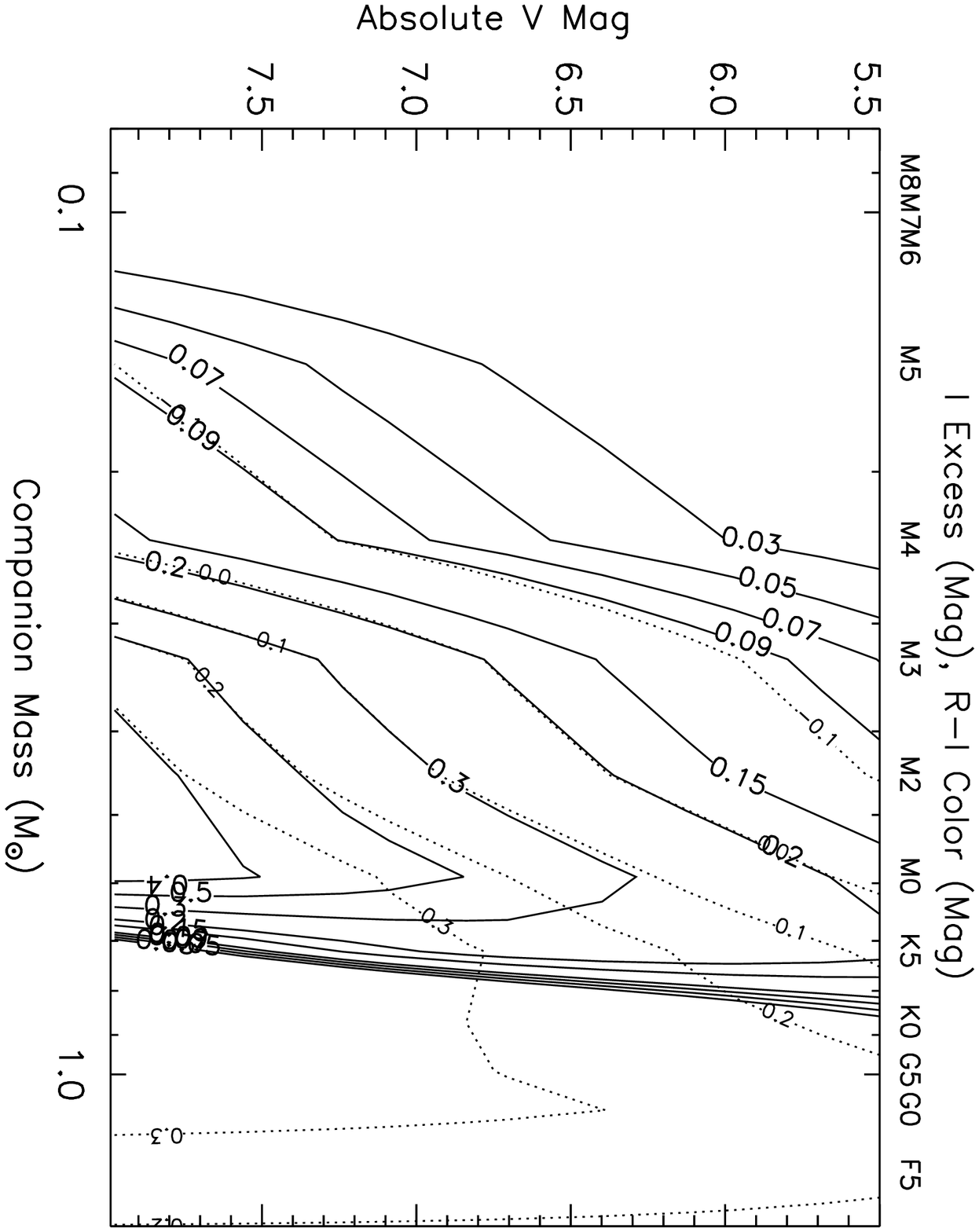} 
		\includegraphics{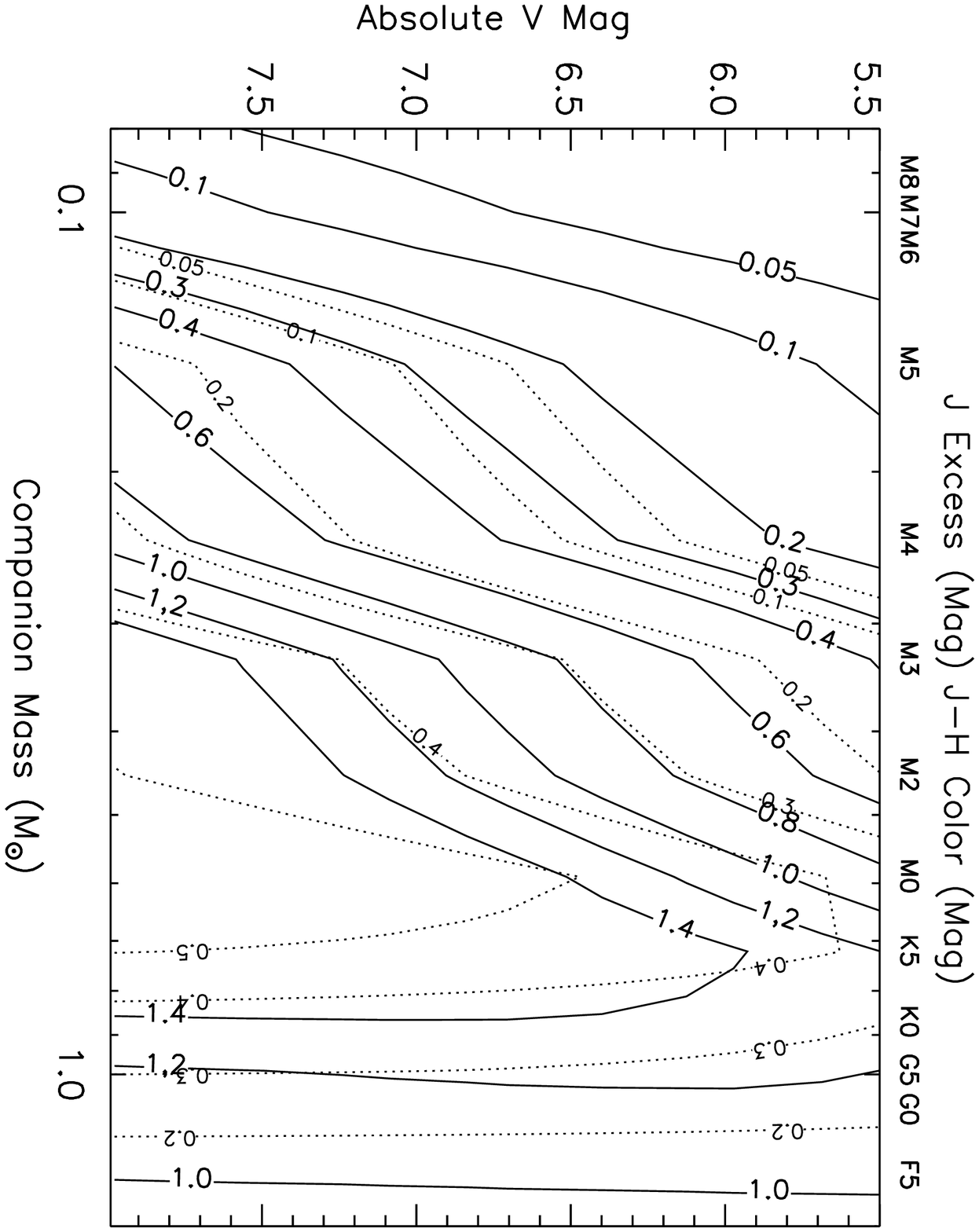}
		\caption{The $I$ and $J$ band excess (solid contours) and $I-R$ and $J-H$ colours (dotted contours) that are produced by main sequence companions (M9V to F5V) as a function of companion mass and primary's absolute $V$ magnitude. These plots are obtained by assuming that the central stars are on the cooling-curve part of the HR diagram and cannot be used to determine the spectral type of the companion of objects which have not yet turned the ``knee" on the post-AGB evolutionary track. The $I$ and $J$ band excess correspond to the $\Delta (V-I)$ and $\Delta(V-J)$ listed in Tables~\ref{tab:resultsI} and \ref{tab:resultsJ}}
		\label{fig:excesses}
	\end{center}
\end{figure*}

To further exemplify the technique, we present in Fig.~\ref{fig:excesses}  the predicted $I$- and $J$-band excess as a function of primary star absolute $V$ magnitude ($M_V$) and main sequence companion mass/spectral type. This plot gives a good idea of the ability of the $I$- and $J$-band methods to detect companions, although it must be pointed out that this figure is only valid for those stars that have already entered the cooling track. To generate Fig.~\ref{fig:excesses}, we created a grid of $M_V$ absolute stellar brightness for the hot central stars using the effective temperatures to determine both bolometric luminosities and bolometric corrections. The former was determined using the average of the cooling track  temperature-luminosity relations of \citet{Schonberner1993}, \citet{Vassiliadis1994} and \citet{Bloecker1995} (for a 1.5~\msun \ main sequence star, but the scatter is very small). The latter was determined using the average of the relation determined by \citet{Vacca1996} and a blackbody curve. To each hot central star we added the flux of a range of main sequence companions using magnitudes and colours reported in Appendix~\ref{app:coolstars}. We thus created a grid of binaries. Next we calculated the total (binary) $B-V$ colour for each primary-secondary combination and compared it to that of the primary alone. A difference is only present when the companion contributes flux in the $V$ or even $B$ and $V$ bands, which is the case for companions brighter than K0-5V. Any difference is interpreted as reddening and used to deredden all bands. We then subtract the primary $I$ band flux from the binary (dereddened) one and any difference is labelled as excess (this is the value reported on each contour line). 

As can be seen in Fig.~\ref{fig:excesses}, a measured $I$ or $J$ band excess corresponds to two distinct companion spectral types. This is due to the contribution of the brightest companions to the $V$ and even $B$ bands, resulting in too high a predicted reddening, too blue a de-reddened spectral energy distribution (SED) and therefore smaller $I$ or $J$-band excess. This effect is less pronounced for the $J$ band, because reddening effects are smaller in that band. The $R-I$ colour provides an approximate way to distinguish between the two companion spectral types allowed by a given $I$ band excess, while the $J-H$ colour is not as discriminating (see dashed contours in Fig.~\ref{fig:excesses}).  
\begin{table}
\begin{tabular}{lcc}
\hline
Band & $\lambda_0$ & $A_{\lambda} / E(B-V)$\\
\hline
$U$ & 3597 \AA & 4.86 \\
$B$ & 4386 \AA &  4.12 \\
$V$ & 5491 \AA &  3.10 \\
$R$ & 6500 \AA &  2.10 \\
$I$   & 7884 \AA & 1.90 \\
$J$  & 1.237 $\mu$m & 0.889 \\
$H$ & 1.645 $\mu$m & 0.562 \\
$K$ & 2.212 $\mu$m & 0.349 \\
$J_{\rm 2MASS}$ &1.241~$\mu$m  & 0.885 \\
$H_{\rm 2MASS}$ &1.651~$\mu$m &0.349 \\
$K_{\rm 2MASS}$ &2.165~$\mu$m & 0.361 \\
\hline
\end{tabular}
\caption{Bandpass central wavelengths after convolution with a 100~kK, logg=7, solar abundance synthetic stellar atmosphere and resulting extinctions according to \citet{Cardelli1989} \label{tab:extinctions}}
\end{table}

In actuality we do not rely on the diagrams in Figs.~\ref{fig:excesses} to determine the companions' spectral types and their limits. Rather we predict the primary's $I$ ($J$) band absolute magnitude using the distance (Table~\ref{tab:data}) and, together with the observed one we derived the secondary's absolute $I$ ($J$) magnitude, which we then convert into a spectral type using the table in Appendix~\ref{app:coolstars}. In this way we do not rely on the $T_{\rm eff}-M_V$ relation characteristic of a given cooling curve. However, the companion spectral types listed in Tables~\ref{tab:resultsI} and \ref{tab:resultsJ} are consistent with those that would be derived using Fig.~\ref{fig:excesses} within one spectral subtype, except for the two objects with companions brighter than spectral type K0V, of which we speak more in \S\S~\ref{ssec:A57} and \ref{ssec:A79}.

\begin{table*}
\begin{center}
	{\begin{tabular}{lcccccc}
	\hline
Name & $E(B-V)$& $(V-I)_0$ & $(R-I)_0$& $\Delta(V-I)$ & $M_{I2}$ & Comp. spec. type \\	\hline
       A~7 &  0.02 $\pm$ 0.02 &  -0.35 $\pm$  0.08 &  -0.19 $\pm$  0.06 &  -0.01 $\pm$  0.08 & $>$  9.60& Later than   M4V    \\
      A~16 &  0.13 $\pm$ 0.03 &  -0.12 $\pm$  0.11 &   0.02 $\pm$  0.09 &   0.22 $\pm$  0.11  &  8.30  [  9.24 --  7.73  ]    &   M3V  [  M3V --  M1V ] \\
      A~20 &  0.07 $\pm$ 0.02 &  -0.32 $\pm$  0.08 &  -0.16 $\pm$  0.06 &   0.02 $\pm$  0.08 & $>$  7.29& Later than   M0V    \\
      A~28 &  0.02 $\pm$ 0.07 &  -0.35 $\pm$  0.24 &  -0.19 $\pm$  0.18 &  -0.02 $\pm$  0.24 & $>$  8.18& Later than   M2V   \\
      A~31 &  0.00 $\pm$ 0.02 &  -0.29 $\pm$  0.08 &  -0.14 $\pm$  0.06 &   0.06 $\pm$  0.08 & $>$  9.14& Later than   M3V    \\
      A~57 &  0.46 $\pm$ 0.04 &  -0.03 $\pm$  0.15 &   0.15 $\pm$  0.12 &   0.29 $\pm$  0.15  &  5.51  [  6.56 --  4.79  ]    &   K3V  [  K7V --  G8V ] \\
      A~71 &  0.38 $\pm$ 0.02 &  -0.31 $\pm$  0.08 &  -0.01 $\pm$  0.07 &   0.05 $\pm$  0.08 & $>$ 10.56& Later than   M5V    \\
       A~72 &  0.04 $\pm$ 0.04 &  -0.35 $\pm$  0.14 &  -0.15 $\pm$  0.14 &   0.01 $\pm$  0.14 & $>$  6.84& Later than   K8V    \\
      A~79 &  1.20 $\pm$ 0.01 &  -0.21 $\pm$  0.10 &   0.42 $\pm$  0.10 &   0.15 $\pm$  0.10  &  3.12  [  4.44 --  2.66  ]    &   F6V  [  G6V --  F3V ] \\
      A~84 &  0.11 $\pm$ 0.03 &  -0.22 $\pm$  0.10 &  -0.08 $\pm$  0.08 &   0.12 $\pm$  0.10  & 10.00  [ 11.91 --  9.28  ]    &   M4V  [  M6V --  M3V ] \\
    DeHt~5 &  0.07 $\pm$ 0.03 &  -0.22 $\pm$  0.11 &  -0.08 $\pm$  0.09 &   0.10 $\pm$  0.11 & $>$  9.28& Later than   M3V    \\
     EGB~1 &  0.20 $\pm$ 0.01 &  -0.29 $\pm$  0.06 &  -0.07 $\pm$  0.05 &   0.07 $\pm$  0.06  & 10.05  [ 11.96 --  8.91  ]    &   M4V  [  M6V --  M3V ] \\
     EGB~6 &  0.03 $\pm$ 0.01 &  -0.34 $\pm$  0.03 &  -0.17 $\pm$  0.03 &   0.01 $\pm$  0.03 & $>$  9.79& Later than   M4V    \\
    HaWe~5 &  0.14 $\pm$ 0.03 &  -0.25 $\pm$  0.10 &  -0.08 $\pm$  0.08 &   0.06 $\pm$  0.10 & $>$ 11.21& Later than   M5V    \\
     HDW~3 &  0.23 $\pm$ 0.01 &  -0.32 $\pm$  0.04 &  -0.06 $\pm$  0.05 &   0.03 $\pm$  0.04 & $>$ 10.35& Later than   M4V    \\
     HDW~4 &  0.04 $\pm$ 0.02 &  -0.24 $\pm$  0.07 &  -0.11 $\pm$  0.06 &   0.07 $\pm$  0.07  & 13.05  [ 16.87 -- 12.30  ]    &   M7V  [  M8V --  M6V ] \\
    IsWe~1 &  0.19 $\pm$ 0.03 &  -0.35 $\pm$  0.09 &  -0.11 $\pm$  0.08 &   0.01 $\pm$  0.10 & $>$  9.77& Later than   M4V    \\
    IsWe~2 &  0.32 $\pm$ 0.04 &  -0.32 $\pm$  0.16 &  -0.04 $\pm$  0.12 &   0.04 $\pm$  0.16 & $>$  9.38& Later than   M4V    \\
       JnEr~1 &  0.00 $\pm$ 0.01 &  -0.37 $\pm$  0.06 &  -0.21 $\pm$  0.05 &  -0.01 $\pm$  0.06 & $>$ 10.86& Later than   M5V    \\
    K~1-13 &  0.00 $\pm$ 0.07 &  -0.42 $\pm$  0.24 &  -0.25 $\pm$  0.19 &  -0.09 $\pm$  0.24 & $>$  8.85& Later than   M3V    \\
     K~2-2 &  0.02 $\pm$ 0.02 &  -0.31 $\pm$  0.09 &  -0.17 $\pm$  0.07 &   0.01 $\pm$  0.09 & $>$  7.40& Later than   M1V    \\
  NGC~3587 &  0.00 $\pm$ 0.04 &  -0.42 $\pm$  0.15 &  -0.23 $\pm$  0.12 &  -0.07 $\pm$  0.15 & $>$  8.96& Later than   M3V    \\
  NGC~6720 &  0.00 $\pm$ 0.03 &  -0.29 $\pm$  0.11 &  -0.16 $\pm$  0.08 &   0.06 $\pm$  0.11 & $>$  7.99& Later than   M2V    \\
  NGC~6853 &  0.00 $\pm$ 0.03 &  -0.32 $\pm$  0.11 &  -0.16 $\pm$  0.08 &   0.04 $\pm$  0.11 & $>$  8.42& Later than   M3V    \\
    PuWe~1 &  0.08 $\pm$ 0.02 &  -0.34 $\pm$  0.09 &  -0.14 $\pm$  0.07 &   0.01 $\pm$  0.09 & $>$ 10.46& Later than   M4V    \\
     Sh~2-78 &  0.31 $\pm$ 0.01 &  -0.26 $\pm$  0.06 &   0.00 $\pm$  0.06 &   0.10 $\pm$  0.06  & 10.22  [ 11.32 --  9.72  ]    &   M4V [  M5V --  M4V ] \\
     Sh~2-176 &  0.26 $\pm$ 0.09 &  -0.30 $\pm$  0.32 &  -0.03 $\pm$  0.25 &   0.05 $\pm$  0.32 & $>$  8.91& Later than   M3    \\
  Sh~2-176$^1$ &  0.23 $\pm$ 0.02 &  -0.25 $\pm$  0.09 &  -0.01 $\pm$  0.07 &   0.11 $\pm$  0.09  & 10.68  [ 12.87 --  9.95  ]    &   M5V  [  M6V --  M4V ] \\
  Sh~2-188 &  0.31 $\pm$ 0.01 &  -0.30 $\pm$  0.05 &  -0.04 $\pm$  0.04 &   0.05 $\pm$  0.05  & 10.46  [ 14.33 --  9.71  ]    &   M4V  [  M8V --  M4V ] \\
     Ton~320 &  0.00 $\pm$ 0.04 &  -0.38 $\pm$  0.14 &  -0.21 $\pm$  0.11 &  -0.05 $\pm$  0.14 & $>$  9.24& Later than   M3V    \\
    WeDe~1 &  0.06 $\pm$ 0.01 &  -0.34 $\pm$  0.03 &  -0.16 $\pm$  0.03 &   0.01 $\pm$  0.03 & $>$ 11.36& Later than   M5V    \\
           \hline
    \multicolumn{7}{l}{$^1$These measurements are obtained by excluding the observations taken in night 3}
\end{tabular}}
	\caption{$I$-band excesses ($\Delta$ ($V-I$)), companion absolute $I$-band magnitudes ($M_{I2}$) and spectral types (or limits) of our targets \label{tab:resultsI}}
\end{center}
\end{table*}

\begin{table*}
\begin{center}
	{\begin{tabular}{lcccccc}
	\hline
	Name & $E(B-V)$& $(V-J)_0$ & $(J-H)_0$& $\Delta(V-J)$ & $M_{J2}$ & Comp. spec. type \\	
	\hline
       A~7 &  0.02 $\pm$ 0.02 &  -0.66 $\pm$  0.11 &  -0.07 $\pm$  0.20 &   0.15 $\pm$  0.11  &  9.65  [ 11.07 --  8.95  ]    &   M5V  [  M8V --  M5V ] \\
      A~31 &  0.00 $\pm$ 0.02 &  -0.41 $\pm$  0.07 &   0.15 $\pm$  0.03 &   0.40 $\pm$  0.07  &  8.26  [  8.57 --  7.98  ]    &   M4V  [  M4V --  M4V ] \\
         A~72 &  0.04 $\pm$ 0.04 &  -0.68 $\pm$  0.17 &  -- &   0.16 $\pm$  0.17 & $>$  6.57& Later than   M1    \\
      A~79 &  1.20 $\pm$ 0.01 &  -0.74 $\pm$  0.06 &   0.02 $\pm$  0.07 &   0.10 $\pm$  0.06  &  4.05  [  5.17 --  3.45  ]    &   G7V  [  K5V --  G1V ] \\
    DeHt~5 &  0.07 $\pm$ 0.03 &  -0.24 $\pm$  0.14 &  -- &   0.54 $\pm$  0.14  &  8.80  [  9.30 --  8.37  ]    &   M5V  [  M5V --  M4V ] \\
     EGB~1 &  0.20 $\pm$ 0.01 &  -0.65 $\pm$  0.11 &  -- &   0.18 $\pm$  0.11  &  9.43  [ 10.52 --  8.80  ]    &   M5V  [  M6V --  M5V ] \\
     EGB~6 &  0.03 $\pm$ 0.01 &  -0.54 $\pm$  0.09 &   0.38 $\pm$  0.10 &   0.28 $\pm$  0.09  &  9.15  [  9.67 --  8.76  ]    &   M5V  [  M5V --  M5V ] \\
        K~2-2 &  0.02 $\pm$ 0.02 &  -0.71 $\pm$  0.09 &  -0.06 $\pm$  0.08 &   0.01 $\pm$  0.09 & $>$  7.84& Later than   M4V    \\
  NGC~3587 &  0.00 $\pm$ 0.04 &  -0.93 $\pm$  0.18 &  -- &  -0.12 $\pm$  0.18 & $>$ 10.12& Later than   M6V   \\
  NGC~6720 &  0.00 $\pm$ 0.03 &  -0.63 $\pm$  0.22 &  -- &   0.19 $\pm$  0.22 & $>$  7.97& Later than   M4V    \\
  NGC~6853 &  0.00 $\pm$ 0.03 &  -0.66 $\pm$  0.11 &   0.05 $\pm$  0.06 &   0.17 $\pm$  0.11  &  8.81  [ 10.05 --  8.12  ]    &   M5V  [  M6V --  M4V ] \\
      Ton~320 &  0.00 $\pm$ 0.04 &  -0.87 $\pm$  0.20 &  -- &  -0.08 $\pm$  0.20 & $>$  9.92& Later than   M6V   \\
       \hline
\end{tabular}}
	\caption{$J$-band excesses ($\Delta$ ($V-J$)), companion absolute $J$-band magnitudes ($M_{J2}$) and companion spectral types (or limits) of our targets. All detections and limits are consistent with the results of the $I$-band excess (Table~\ref{tab:resultsI}) \label{tab:resultsJ}}
\end{center}
\end{table*}

\begin{figure*}
\vspace{9.5cm}
	\begin{center}
		\includegraphics{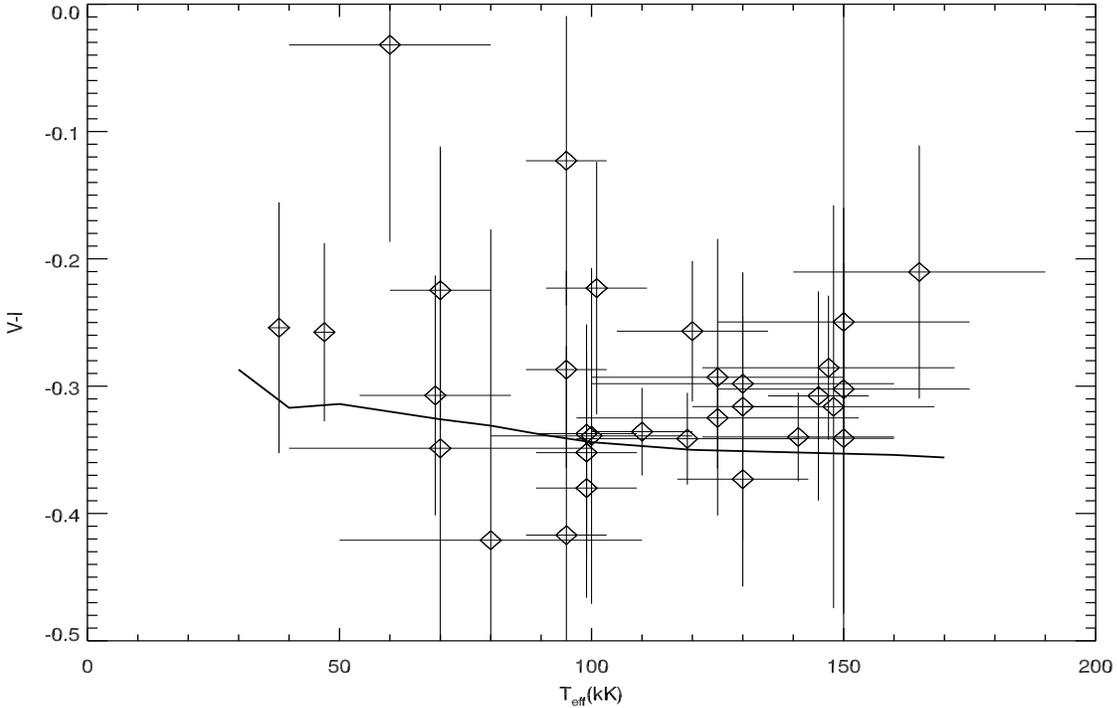} 
		\caption{The observed (dereddened) $V-I$ colours of our targets (symbols; see Table~\ref{tab:resultsI}) compared with the predicted $V-I$ colours of single stars as a function of effective temperature (for a $\log g = 7.0$ -- solid line) }
		\label{fig:resultsI}
	\end{center}
\end{figure*}

\begin{figure*}
\vspace{8cm}
	\begin{center}
		\includegraphics{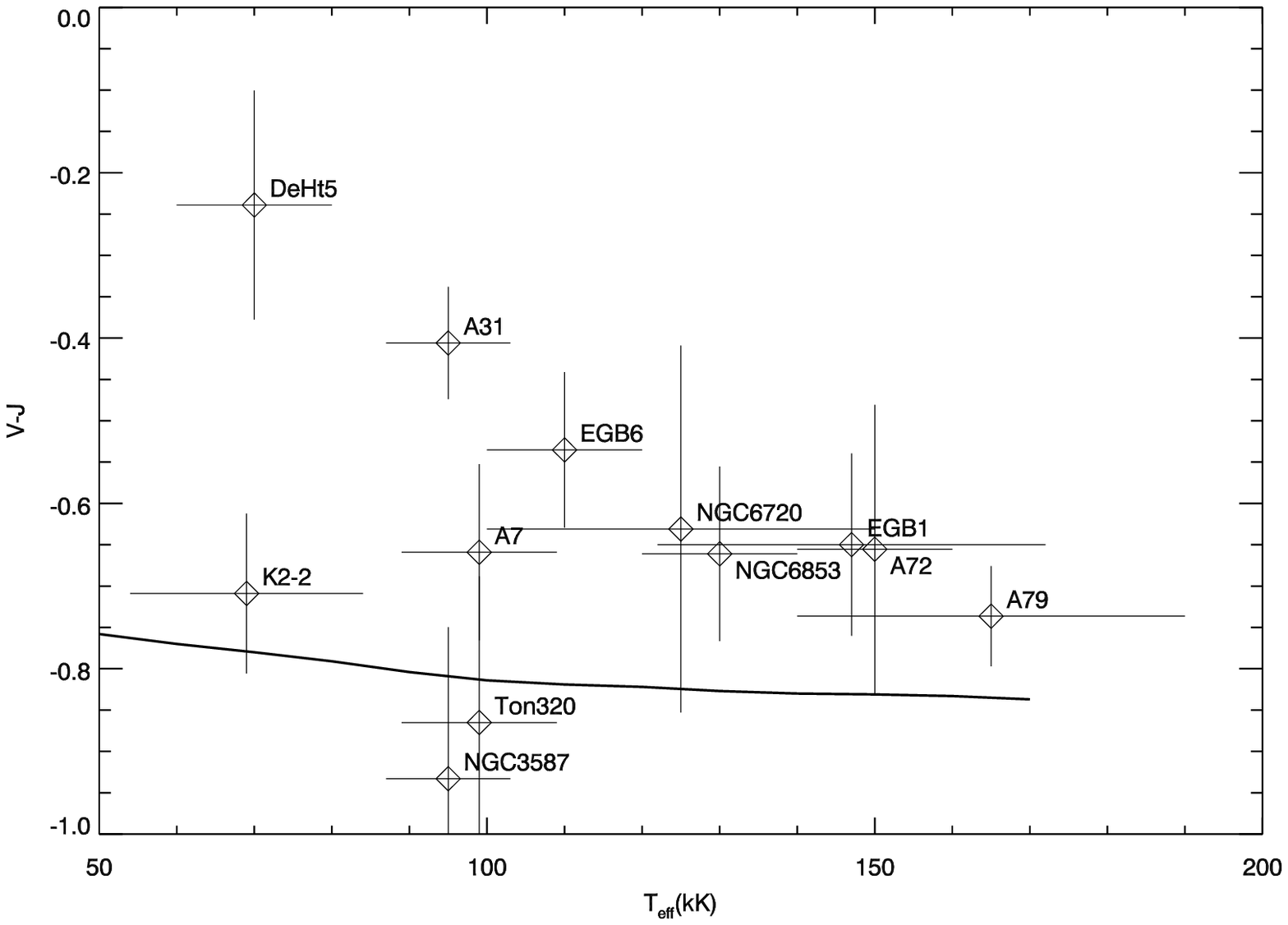} 
		\caption{The observed (dereddened) $V-J$ colours of our targets (symbols; see Table~\ref{tab:resultsJ}) compared with the predicted $V-J$ colours of stars as a function of effective temperature (for a $\log g = 7.0$) }
		\label{fig:resultsJ}
	\end{center}
\end{figure*}


\section{Results}
\label{sec:results}

Those objects for which $\Delta(V-I) > \sigma_{\Delta(V-I)}$ (or, equivalently, $\Delta(V-J) > \sigma_{\Delta(V-J)}$) are considered as cases where a companion is detected. We note that this difference is the same as what we call the $I$ ($J$) band excess in Fig.~\ref{fig:excesses}, because the theoretical colours are normalised to the observed (dereddened) $V$ magnitudes. Reddenings, intrinsic $V-I$ ($V-J$) and $R-I$ ($J-H$) colours, $I$ ($J$) band excesses ($\Delta(V-I)$ or $\Delta(V-J)$), companions' absolute $I$ ($J$) band magnitudes and companions' spectral types are listed in Tables~\ref{tab:resultsI} (\ref{tab:resultsJ}). The $I$ ($J$) band excesses are plotted as a function of stellar temperature in Fig.~\ref{fig:resultsI} (\ref{fig:resultsJ}).   

\begin{figure}
\vspace{8cm}
	\begin{center}
		\includegraphics{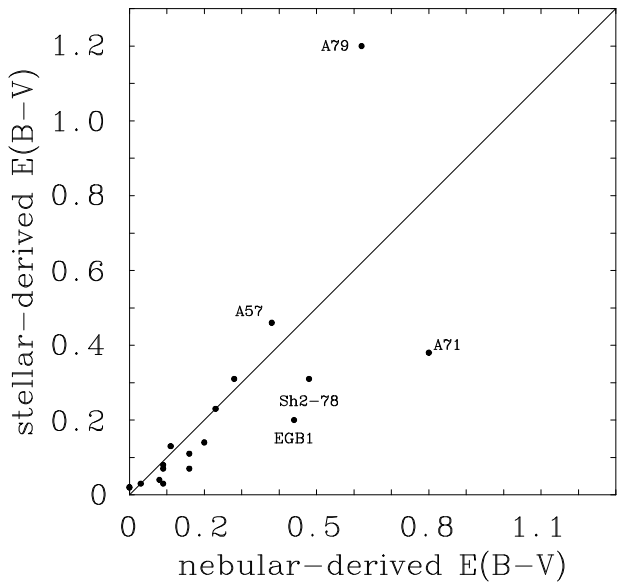} 
		\caption{A comparison of the reddening values obtained via a nebular method (Table~\ref{tab:data}), and those obtained via our stellar method (Table~\ref{tab:resultsI})}
		\label{fig:reddeningcomparison}
	\end{center}
\end{figure}

The reddenings derived by comparing observed and (single star) predicted $B-V$ colours are compared to those derived from nebular methods in Fig.~\ref{fig:reddeningcomparison}. Most reddenings compare well. The five exceptions are A~57 and A~79, whose high derived stellar reddenings are likely due to bright companions affecting the $B$ and $V$ bands, A~71, EGB~1 and Sh~2-78, for which we suspect the nebular reddenings have uncertain values.

We have 9 detections or marginal detections in the $I$ band: 
A~16 (at the 2-$\sigma$ level), 
A~57 (at the 1.9-$\sigma$ level), 
A~79 (at the 1.5-$\sigma$ level), 
A~84 (at the 1.2-$\sigma$ level), 
EGB~1 (at the 1.2-$\sigma$ level), 
HDW~4 (at the 1-$\sigma$ level, but this is a PN mimic),  
Sh~2-78 (at the 1.7-$\sigma$ level), 
Sh~2-176 (at the 1.2-$\sigma$ level, only for the measurement excluding night 3) and
Sh~2-188 (at the 1-$\sigma$ level). 
A~79 was already known to have a F0V companion \citep{Rodriguez2001}. The spectral type determined with our method is cooler because the reddening was overestimated due to the strong contribution of the companion in the $V$ band (see \S~\ref{sec:technique}). We have therefore 8 detections out of 27 {\it bona fide} objects.

In the $J$ band we detected 
A~7 (at the 1.4-$\sigma$ level), 
A~31 (at the 4.7-$\sigma$ level), 
A~79 (at the 1.7-$\sigma$ level), 
DeHt~5 (at the 3.8-$\sigma$ level, but this is a PN mimic), 
EGB~1 (at the 1.6-$\sigma$ level), 
EGB~6 (at the 3.1-$\sigma$ level) and 
NGC~6853 (at the 1.5-$\sigma$ level). We have therefore 6 detections out of 11 {\it bona fide} central stars with data.

The spectral types of the companions implied by the detected excesses as well as limits of the non-detections are also listed in Tables~\ref{tab:resultsI} and \ref{tab:resultsJ}. When a companion was not detected, we summed the $I$ ($J$) band excess with the upper error bars to create an upper limit to the flux excess and determined an upper limit for the companion mass/spectral type in this way. We note that all limits and detections are fully consistent across the two detection techniques.

The $I$ band binary fraction is determined by the ratio of 8 objects over 27 {\it bona fide} central stars (see \S~\ref{sec:sample} for an explanation of what we mean by {\it bona fide}), or 30\%. Only companions brighter than the spectral type M3-4V can be detected by our survey, where this limit was estimated by taking the median of the limits in Table~\ref{tab:resultsI}. Within the $J$ band group, the detection rate is 6 out of 11, or 54\%, in line with the study of \citet[][see also Frew 2008]{Frew2007}. A look at Table~\ref{tab:resultsJ} shows that most detections in the $J$ band were fainter, in line with the expectation that the $J$ band is a more sensitive method. Based on the detection and limits of Table~\ref{tab:resultsJ} we guesstimate that this limit is M5-6V. The binary fraction for the entire sample, obtained from either the $I$ or $J$ band excess methods is 44\% (12 out of 27 bona fide PN), to a limit intermediate between the two methods. 

Before comparing our results with predictions, we would like to remark on the asymmetry of the distribution of $V-I$ colours of our sample about the single star prediction line (Fig.~\ref{fig:resultsI}). Even eliminating from that plot the 9 detections, which are all above the prediction, the non-detections (21 objects) still preferentially lie above the single star prediction line (15 vs. 6 objects). We speculate that this points to the detection of a binary signal which is higher than the fraction we determined from individual objects above. To determine whether this is significant we would need a Monte Carlo simulation. One could counter argue that this asymmetry is caused by an incorrect placement of the prediction line, or, in other words, that the predicted colours are too blue. To shift the prediction line so that the non detected data points scatter symmetrically about it, would require a systematic shift of the synthetic $V-I$ colours of about $\sim$5\% which is excluded by the comparison of synthetic spectra with data carried out over the years (Rauch, private communication; \citealt{Rauch2007}).

\begin{figure*}
\vspace{6cm}
	\begin{center}
		\includegraphics{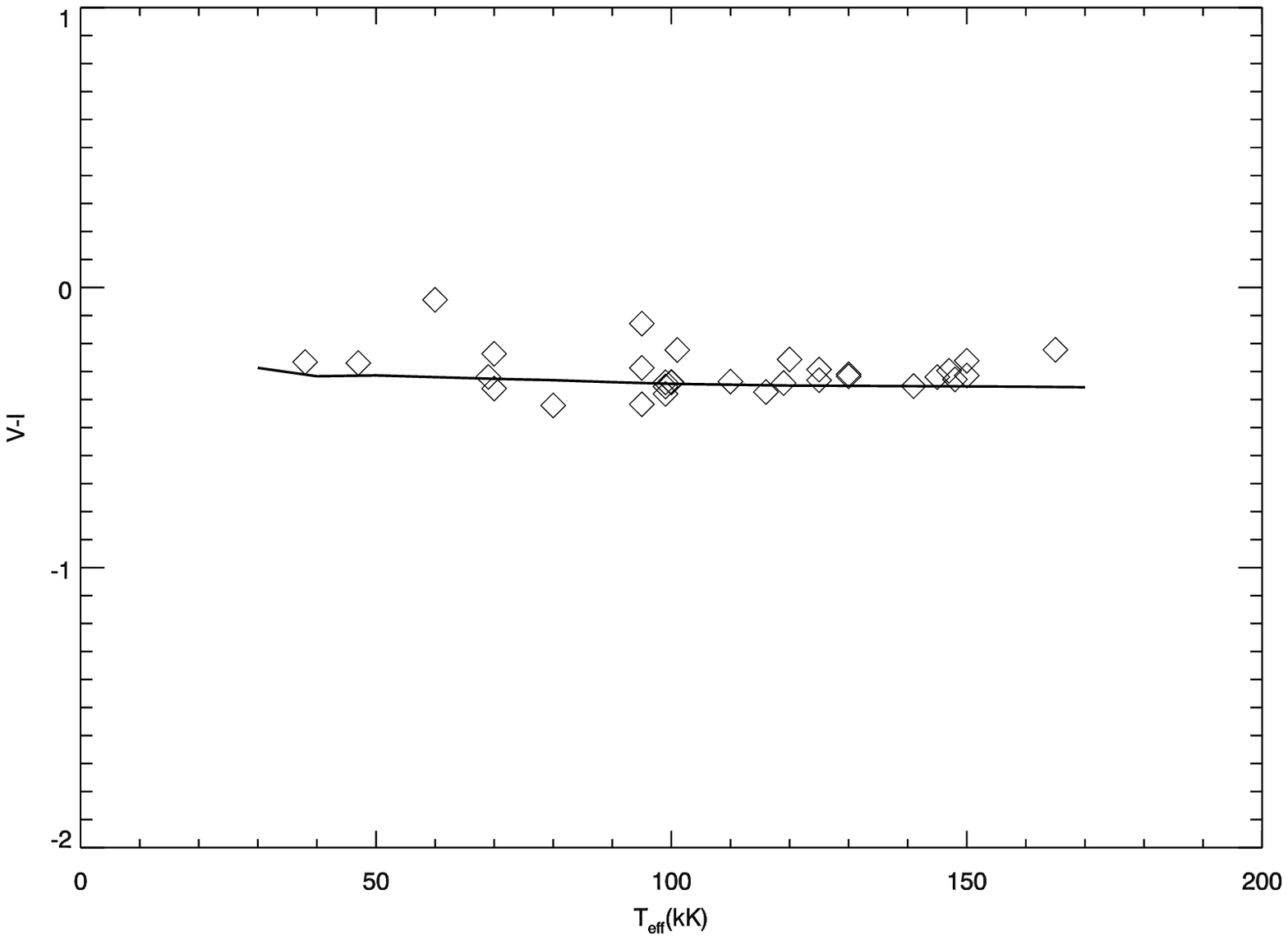}
		\includegraphics{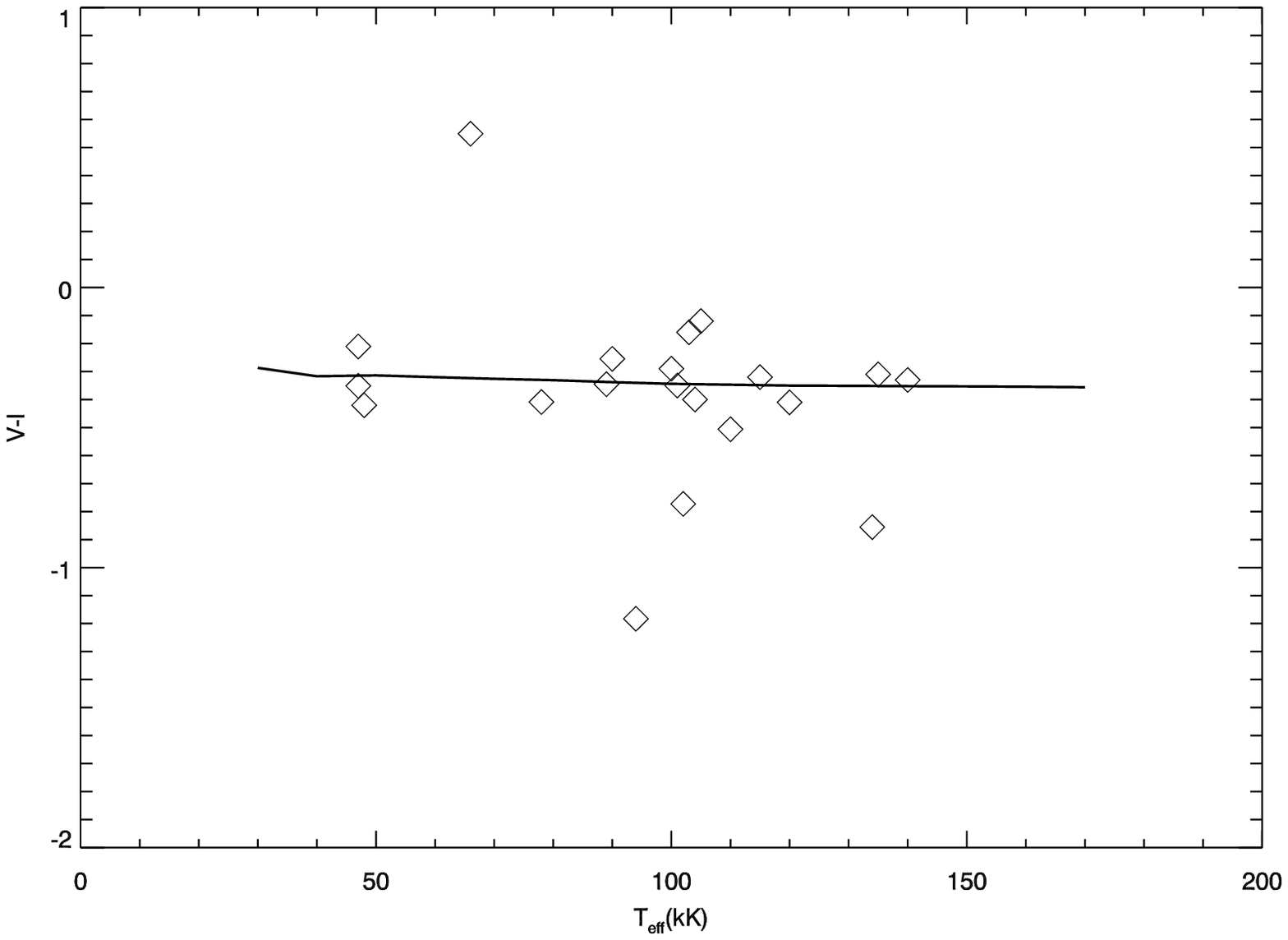} 
		\caption{Left panel: the same figure as Fig.~\ref{fig:resultsI}, except here we compare our results (with no error bars, for clarity) with results obtained using the data compiled by \citet[][right panel]{Bilikova2012}, showing how unvetted data has much larger random errors}
		\label{fig:Bilikova}
	\end{center}
\end{figure*}

Finally, we have demonstrated that the $I$ band excess method necessitates extraordinary accuracy, such as that achieved for the current dataset, but is not usually encountered in unvetted data from the literature. For example, in Fig.~\ref{fig:Bilikova} we compare the $I$-band excesses obtained by using the $B$, $V$ and $I$ data compilation of  \citet{Bilikova2012} with our dataset. Using all of their data and determining the stellar temperatures in the same way we have done for our sample, we see how the scatter around the single star prediction is dramatic. While their compilation was not aimed at detecting binarity, it can easily be seen that data from the literature is not generally suitable for this type of work.

\subsection{Photometric Variability}
\label{ssec:variability}

Periodic variability in central stars of PN usually denotes a short period binary \citep[e.g.,][]{Miszalski2009}. Variability should not interfere with the detection of an IR excess because the exposures in each filter are taken in a short sequence and the periods of these objects are of the order of 12 hours \citep{DeMarco2009}. Although it is expected that the colour of the short period binary should vary during the orbit, this is only marginally so \citep[e.g.,][]{DeMarco2008c}.  Variability can however increase the error on the absolute photometry because we average observations taken at more than one epoch. We have therefore monitored all targets for variability during photometric as well as non-photometric nights. The differential magnitudes to 5 field stars are then plotted as a function of time. The central stars for which the scatter of the average differential magnitudes in each filter is larger than the mean errors on these magnitudes are labelled as possible variables. Further, we retain as likely variables those of these stars for which the variability trends are the same in all filters. We will continue monitoring these stars to determine if a period is present. The variable stars are: A~57, A~72, A~84, IsWe~2, Sh~2-78 and Sh~2-176. Four of the 6 variables have an $I$ band excess, as one might expect if the variability denotes binarity. Further comments on the variables are reported in \S~\ref{sec:individualobjects}.

\section{Comparison of the overall PN binary fraction with the overall main sequence binary fraction}
\label{sec:prediction}

We recognise that the binary fraction determined here is preliminary, because of the small sample size. However, here we compare it with the prediction from the current PN evolution scenario, whereby PN derive from {\it all} $\sim$1-8-\msun\ stars, whether they are single stars or are in binaries. As already explained, it is this prediction that we are testing. We expect that if PN derive from a binary interaction more often than for the current evolutionary scenario, we would find a larger fraction of short- and intermediate-period binaries in the central star population. This leads to the prediction that the {\it overall} binary fraction, i.e., where we consider binaries at any period, should also be larger in the {\it binary hypothesis} than in the standard scenario.

The progenitor population of the PN from which we have drawn our targets is the main sequence stars with masses between $\sim$0.9 and 8~\msun. We can then use the main sequence population binary fraction of \citet{Raghavan2010} to represent the binary fraction of the progenitor population of our PN. We use (50$\pm$4)\%, which is the percentage of double and multiple star systems with primaries in the spectral type range F6V-G2V (masses in the range 1.27 -- 1.03~\msun; see Table~\ref{tab:mainsequencecolours}). This is reasonable on the grounds that the median progenitor mass of today's PN is 1.2~\msun\ \citep{Moe2006}. This fraction is lower than that determined by \citet[][57\%]{Duquennoy1991} who accounted for a larger incompletion bias, but is likely to be more accurate. This fraction includes {\it any main sequence companion down to the planetary regime at any separation}.  

\subsection{Accounting for completion effects}
\label{sssec:completion}

Before we can carry out a comparison we need to increase the PN binary fraction to (i) include the companions that have not been detected because they are too faint; (ii) include wide companions, because our observed sample does not contain resolved binaries by design. Finally, we need to (iii) account for the fact that some main sequence close binaries go through a common envelope interaction on the RGB. Those systems will become short period binaries with very low mass envelopes (the primary will become a subdwarf O or B star; \citealt{MoralesRueda2003}) and are unlikely to ever ascend the AGB on the grounds of low envelope masses \citep{Dorman1993}; if they do, they will suffer a second common envelope almost immediately which would result in a sub-luminous object, highly unlikely to ever make a visible PN.

(i) To account for companions with an M3-4V spectral type or fainter (mass $\le$0.33-0.24~\msun), we use Table~18 of \citet{Raghavan2010} that lists all the companion spectral types of the detected binaries. We determine that the fraction of main sequence binaries with a companion spectral type of M3-4V or later is 41-31\% (73-56 of 179 main sequence binaries for which the companion spectral type is known, where we count as undetectable also 4 hot and evolved companions). We therefore increase our $I$-band derived PN binary fraction (30\%) by a factor of 1.69-1.45 to account for the undetected faint companions.

To account instead for companions with brightness equal to, or fainter than M5-6V (the limit we estimated for the $J$ band dataset) we would have to multiply the $J$-band-derived PN binary fraction (54\%) by a factor of 1.28-1.19.

(ii) The ground-based spatial resolution of our observations is probably only slightly smaller than the median seeing of 1.3~arcsec corresponding to a projected separation of $\sim$1040~AU for the median distance of our sample of 0.80~pc. 
To de-project this separation we would have to divide the projected separation by a factor of 0.8, which accounts for the systems' random phase {\it and} random orientation. However, if we also account for an eccentricity distribution similar to the main sequence's, this factor approaches unity, because systems spend more time near apastron in eccentric systems.  Not knowing the eccentricity distribution of central star binaries but presuming a degree of circularisation to have taken place, we de-project the projected separation using a range of factors (0.8-1.0). The median separation is then converted to a period in the range $\sim$13\,000--18\,000 years (log(P/days)=7.11-7.25), using Kepler's third law and a total system mass of 0.9~\msun. This period limit is then decreased to $\log(P/{\rm days})=6.71-6.85$, to account for the fact that, due to mass-loss and angular momentum conservation, the orbit of a typical binary will widen by a factor of 2-3 (we used 2.5) and so some of the main sequence binaries of \citet{Raghavan2010} will become resolved central star binaries in our sample because of orbit widening. 

A similar argument can be applied to the $J$-band sample of 11 bona fide PN, which have a median distance of 0.65~kpc, with a resulting median projected separation of 845~AU, resulting in a period corrected for orbit widening of $\log(P/{\rm days})=6.57-6.71$.

(iii) Finally, the main sequence binary population with period shorter than log(P/days)=2.43 never ascend the AGB. This value was calculated using a radius on the RGB of 100~\rsun (see, e.g., see figure 5 of \citet{DeMarco2011}), and a maximum tidal capture radius of 2 stellar radii (using a total system mass of 1.5~\msun). The maximum tidal capture radius for the RGB was obtained from \citet{Villaver2009} and \citet{Madappatt2011}.  At this point we integrate under the normalised main sequence binary period distribution of \citet[][their Fig.~13]{Raghavan2010} using log(P/days)=2.43 - 6.78 as limits (or 2.43 -- 6.65 for the $J$-band sample; where we have taken the average of the period separation ranges to determine the upper limits), obtaining a fraction of 0.65 of the total (or 0.64, for the $J$ band sample): these are the systems we detect. This translates into a second factor of 1.54 (or 1.56), which we multiply by the central star binary fraction so as to include binaries at all separations and thus make it comparable to the main sequence one.

(iv) The only bias that is impossible to account for, is that due to central stars whose companions are bright enough to contribute to the $V$ and even $B$ bands. As we have explained in \S~\ref{sec:technique}, these will result in reddenings that are artificially large, because the bright companions tend to make their system's colours redder. Once this over-large reddening is applied to all the bands, the binary SED is rendered bluer than it should be and the $I$ or $J$ band excess is necessarily reduced. The mismatch between reddening curve and SED of the companion can decrease the red/IR flux excess below detectability or reduce it such that it is impossible to match it to a companion spectral type. Interestingly, only 6 objects have stellar-derived reddening values that are larger than the nebular-derived values, the most noticeable being A~79, a known binary and A~57 which has a large $I$ band excess.  For the other 4 objects (A~7, A~16, A~28 and Sh~2-188), these discrepancies are however always within the uncertainties. We therefore conclude that either the reddening comparison is not reliable due to high errors in nebular-derived reddenings, or that our sample tends not to include companions much hotter than spectral type K0V. This makes sense in view of the fact that only a minority of companions are expected to be that hot). We leave this bias unquantified in the knowledge that it would be at most a few percent. A spectroscopic follow-up will resolve this issue.

\subsection{The debiased PN binary fraction and its uncertainties}
\label{sssec:fraction}

The de-biased central star binary fraction obtained through the $I$ band photometry of our sample is 67-78\%. Using the $J$ band data alone we obtain a fraction of  100-107\%.  The ranges account for the uncertainty on the factor to account for the unobservable faint companions. The error bars on these estimates are difficult to estimate for the moment and we defer this exercise to the next paper in this series, which will increase the sample. However, if all limits were accurately accounted for then the $I$- band and $J$-band fractions should be the same. We therefore tentatively estimate a 10-30\% error bar. We preliminarily conclude that the PN binary fraction is higher than the main sequence binary fraction of (50$\pm$4)\%.

We can also estimate the PN binary fraction after eliminating from each of the $I$ and $J$ samples the one detection with the smallest statistical relevance. By doing so, we would obtain $I$ and $J$ binary fractions of  58-68\% and 84-90\% respectively.

\subsection{Comparison of the short-period PN binary fraction with the main sequence binary fraction}
\label{sec:prediction2}

The fraction of PN that surround short-period, post-common envelope binaries is 15-20\%  \citep{Bond2000,Miszalski2009}. This is a very large fraction when we consider that only a very small minority of main sequence binaries would suffer a common envelope on the AGB.

The fraction of main sequence binaries that go through a common envelope on the AGB, resulting in post-common envelope central stars of PN should be quite small: only those companions that escape capture on the RGB, but then are successfully captured on the AGB will become post-common envelope central stars. These are companions that, while on the main sequence, have periods in the range $\log (P/{\rm days}) > 2.7 - 2.8$ and therefore account for only $\sim$1\% of all main sequence binaries. The lower period limit was calculated  in \S~\ref{sec:prediction}, while the higher period limit was obtained by considering that for a successful AGB capture the orbital separation has to be smaller than approximately twice the maximum AGB radius \citep{Villaver2009,Nordhaus2010,Madappatt2011}. The maximum AGB stellar radius for a 1.2-\msun\ star was calculated to be $\sim$300~\rsun\ (\citealt{DeMarco2011}, with the usual adjustment for the orbital widening due to mass loss, see \S~\ref{sec:prediction}). Even exaggerating the maximum AGB radius (600~\rsun) and the maximum capture radius to 5 stellar radii, we get $\log (P/{\rm day}) = 3.9$ and a fraction of 5\%. We would therefore predict a similar fraction of post-common envelope central stars, contrary to the observations.

\subsection{Comparison of the PN binary fraction with the white dwarf binary fraction}
\label{sec:WD}

We note, finally, that the PN binary fraction, even before detection biases are accounted for, appears to be higher than the white dwarf binary fraction of 25\% \citep[][the WD binary fraction should be a few points higher, if we could readily detect white dwarfs around bright main sequence stars - Sirius-like systems]{Holberg2009}. One reason could be that the white dwarfs derive from a population that includes lower mass stars, those that do not develop a PN due to long transition times. Such lower mass population would naturally have a lower binary fraction \citep{Raghavan2010}. Another reason could be that PN form preferentially around binaries, the hypothesis we are trying to test. To discern between the two explanations of the discrepancy between the white dwarf and PN binary fractions we need better constraints on white dwarfs and central stars masses \citep[][]{Liu1995,Liebert2005,Gesicki2007}.


\section{Notes on remarkable individual objects}
\label{sec:individualobjects}

\subsection{Abell 7}
\label{ssec:A7}
This star was detected in the $J$ band (at the 1.4-$\sigma$ level) and the implied companion would be an M5V star. Companions dimmer than M4V are excluded by the $I$ band photometry.  The central star has a red-dwarf companion 0.91 arcseconds away from the central star (Ciardullo et al. 1999) resolved by the {\it Hubble Space Telescope (HST)}, which warrants further investigation in order to determine a photometric parallax; only an upper limit of K2V could be determined from the $HST$ data of Ciardullo et al. (1999), while Frew (2008) determined a spectral type of M4V, similar to our own.  

\subsection{Abell 16}
\label{ssec:A16}
We detected an M3V companion in the $I$ band at the 2-$\sigma$ level.

\subsection{Abell 31}
Ciardullo et al. (1999) give an upper limit to the distance of 440~pc (based on photometry of the resolved companion),  mildly inconsistent within the uncertainties with  a direct trigonometric determination of $D$ =  568$^{+131}_{-90}$ pc \citep{Harris2007}. Our $I$ and $J$ photometry suggest an M4V companion, in agreement with Frew (2008) and \citet{Ciardullo1999}, who used the $HST$ to determine a separation of 0.26 arcsec for a companion with spectral type later than M4V. This companion may never have interacted with this PN. The morphology of this PN is likely round, in agreement with the lack of interaction.

\subsection{Abell 57}
\label{ssec:A57}
The $I$-band excess detected for this object exceeds the uncertainty by a factor of two and the spectral type of the companion is K3V. Such a bright companion may have contributed in the $V$ band. In fact, the reddening derived from our method is higher than the reddening derived from nebular observations. (cf. Tables~\ref{tab:data} and \ref{tab:results}). We also note that we used the limiting temperature of 60kK for this star. A higher temperature would lower the $I$ band excess slightly, with a temperature of 150kK reducing it to 0.25~mag.  \citet{Miszalski2011} discovered this to be another example of a EGB~6-like central star \citet{Frew2010c}, implying the presence of a compact disk either around the central star or its companion, lending strong support to this star being a physical binary.  This central star appears to be variable at the 0.1-mag level from our differential photometry, while our absolute photometry reveals at most a variability of a 2\% based on two epochs of data.

\subsection{Abell 72}
\label{ssec:A72}
As was the case for A~57, the effective temperature of the central star of A~72 can be constrained to be larger than 100~kK by the Zanstra method. We therefore adopt a lower limit of 100~kK due to the fact that our own UVES spectrum suggests a temperature in excess 120~kK. There is effectively no difference in the calculated excess for temperatures higher than 100~kK. From $I$ and $J$ band photometry we can state that the companion, if present, would be fainter than the M1V spectral type. We suspect this star to be variable at the 0.05-0.1-mag level.

\subsection{Abell 79}
\label{ssec:A79}

This star is known to have a cool spectral type \citep{Rodriguez2001} possibly denoting that it is a binary since the ionising source of the PN would otherwise be missing. When the cool companion dominates the light of the system, we can assume that it contaminates the $V$ or even the $B$ bands. As a result, the  reddening will be too large. We determined $E(B-V)=1.2$ while from near-IR data a value of 0.62 is derived (Frew 2008). Using our reddening, all the photometry is de-reddened excessively, with resulting derived $V-I$ and $V-J$ colours that are too blue and a reduced $I$ and $J$ band excess. That is why the determined companion spectral type (F6V from $I$ band photometry and G7V from $J$ band measurements) are redder/fainter than the one that was determined from spectroscopy (G0V) by \citet{Rodriguez2001}. Note how the result from $J$-band photometry is closer to the spectroscopically-derived spectral type as predicted in \S~\ref{sec:technique}. In addition, we cannot use the contour plots in Figs.~\ref{fig:resultsI} and \ref{fig:resultsJ} because we do not have the central star's $M_V$ value.

The $I$ band excess for this star is large (0.15 mag) but it is affected by a considerable uncertainty. The reason is that the $I$ band magnitude of this star is 0.2 mag brighter in night 4 and in night 3. While we could exclude the N3 measurement on the grounds that the night was only partly photometric, our logs show that that part of the night was completely clear. In addition, the magnitudes in all other bands are very similar across the two nights. We therefore keep both measurements and accept the large uncertainty.

\subsection{Abell 84}
\label{ssec:A84}
The central star of A~84 has a companion detected in the $I$ band at the 1.2-$\sigma$ level, making it a marginal detection.
The differential photometry shows a clear dimming trend over the 8 nights of data at the 0.1 mag level.

\subsection{DeHt~5}
\label{ssec:DeHt5}
This large emission nebula (also known as DHW5) was first identified by \citet{Dengel1980} as a likely PN.  It has an unusual irregular  morphology, and shows quite a different appearance in [O~{\sc iii}] light, compared to its morphology in [N~{\sc ii}] and [S~{\sc ii}] emission \citep{Rosado1991,Tweedy1996}.   Previous researchers have assumed that DeHt~5 is a bona fide PN.  \citet{Bannister2003} had previously noted the proximity of its central star, WD~2218-706, to the giant molecular cloud complex described by \citet{Kun1998}, and stated that the star may lie in an area where the ISM is dense.  However, these authors did not consider the possibility that the optical nebula might be in fact ionised ISM.   Tweedy \& Kwitter (1996) also noted the extensive diffuse emission around the bright nebula, suggesting this might be ionised interstellar material, but assumed that the core was a true PN.   Frew (2008) found that the nebular and stellar velocities are different, suggesting that the nebula is unrelated to its ionising star, a situation similar to PHL~932 \citep{Frew2010b}.  The overall body of evidence is again in favour of the ionised-ISM interpretation, rather than a PN.  The morphology  is not typical of an evolved PN,  and the gas is consistent with being ionised ambient material, as the systemic velocity agrees with the CO velocities of widespread neutral gas at $\sim$360 pc, the distance of the star as determined by  Benedict et al. (2009).  The narrow line width of the nebular gas is also consistent with the ionised ISM interpretation.  Furthermore, the evolutionary position of the star in the HR diagram is {\it not} in agreement with a post-AGB track that is consistent with the timescale of PN evolution.  All of the evidence is consistent with DeHt~5 being a HII region ionised by WD~2218+706. The central star has a detected $J$-band excess, consistent with an M3V companion, but we flag that this is one of two objects with one epoch of measurements only.

\subsection{EGB~1}
\label{ssec:EGB1}
This somewhat irregular, one-sided nebulosity is traditionally classified as a PN. It is poorly studied, but spectra show moderately bright [O~{\sc iii}] emission.  However, the lack of a limb-brightened bow shock (expected for such a one-sided nebula with an off-centre central star), the small line-width \citep{Lopez2012}, the systemic nebular velocity close to zero in the local standard of rest (LSR), and the moderate excitation, lead to questions regarding its status.   It may be another case of ionised ISM, but more observations need to be obtained before a definitive answer can be given and therefore we have kept it as a PN for the time being.   Additionally, the ionising star is a periodic photometric variable (T. Hillwig, private communication), consistent with binarity.  We detected an $I$ and $J$ band excess at the 1.2 and 1.6-$\sigma$ levels, respectively, consistent with an M4-5V companion. The hypothesis that the companion to this central star could be a white dwarf is inconsistent with the $R-I$ colour.

\subsection{EGB~6}   
\label{ssec:EGB6}
\citet{Liebert1989} found an unresolved, dense  emission nebula surrounding the central star, with a tiny ionised mass of $\sim$10$^{-9}$~M$_{\odot}$.   \citet{Zuckerman1991} found near-IR evidence for a probable red dwarf companion, confirmed by \citet{Fulbright1993} via $JHK$ photometry of the central star.    The companion has been imaged with $HST$ (see \citealt{Bond1993} and \citealt{Bond1994}) at a projected separation of 0.18\arcsec.  Furthermore, the unresolved emission nebula corresponds in position to the red dwarf and not the central star \citep{Bond1994}, suggesting there may be a small disk of material around the cool companion (possibly accreted from the PN), and photoionised by the central star.  This is the archetype of a class of PN central stars with compact, unresolved ionised nebulae (or disks), which \citet{Frew2010c} refer to as EGB~6-like central stars. 
We detected a 3-$\sigma$ $J$ band excess consistent with an M5V companion, in line with the results of \citet{Bond1993}, but we note that the $J-H$ colour is mildly inconsistent with this spectral type, as it appears too red.

\subsection{HDW 4}
Also designated HaWe 6, this faint nebula was noted by \citet{Hartl1983}.   \citet{Harris2007} determined a trigonometric distance to the ionising star of only  209$^{+19}_{-16}$ pc.    Napiwotzki (1999) performed a NLTE model atmosphere analysis, determining $T_{\rm eff}$ = 47,300 $\pm$ 1700 K and log\,$g$ = 7.93 $\pm$ 0.16, and estimating a distance of 250\,pc.  The relatively low temperature and high gravity of the white dwarf suggests a long cooling age far greater than any feasible PN lifetime.  Referring to Table~1 of \citet{Bergeron1995}, a cooling age of 3--4$\times10^{6}$~years is suggested.  The absolute magnitude is $M_{V}$ = 9.43$^{+0.18}_{-0.19}$ which is considerably fainter than any PN nucleus.  The very low ionised mass of the nebula (5 $\times$  10$^{-3}$~$M_{\odot}$; Napiwotzki 1999) also rules out a PN interpretation.  \citet{Napiwotzki1999} speculates that the nebula might be a shell produced by an ancient nova outburst (the mass is approximately right), but the star shows no sign of any cataclysmic variable features in its spectrum \citep{Napiwotzki1995}.  Napiwotzki (1999) obtained a spectrum of the HII region around this star and deduced an upper limit to the expansion of 2$v_{\rm exp}$ $<$ 47 \kms.  Hence an expanding nova shell hypothesis is ruled out.  The expansion limit is consistent with ambient ISM, so the nebula may be another case of ISM ionisation by the unrelated hot white dwarf WD~0533+555 (Frew 2008). We report a very marginal 1-$\sigma$ detection in the $I$ band, consistent with an M7V companion. However, this object was observed only once and that observation may have been affected by thin cirrus (Table~\ref{tab:logs}). 

\subsection{HaWe~5}
For the `central' star, Napiwotzki (1999) determined $T_{\rm eff}$ = 38,100 $\pm$ 1500 K and log\,$g$ = 7.58 $\pm$ 0.20 from an NLTE model atmosphere analysis.  The estimated distance is $\sim$420\,pc.   This small faint nebula shares characteristics in common with HDW~4, where the large evolutionary age of the star is greatly at odds with the existence of a remnant PN.   \citet{Napiwotzki1999} also suggests that this may be an old nova shell and estimated an ionised mass of only 2$\times$ 10$^{-4}$ $M_{\odot}$.  A red {\it Digitized Sky Survey} image shows a vaguely PN-like elliptical nebula surrounding the ionising star.  The small size and faintness of the object would be remarkable if it was a real PN at the nominal distance.  Instead, it is more likely to be be a wisp of ionised ambient material.  We note that the second epoch of this object was suspected non-photometric. The fluxes in the four bands for that exposure were indeed fainter than the other two by 0.03, 0.23 0.15 and 0.01 mag. While clouds are grey and one would expect a similar magnitude drop in all bands, it is possible that the high cirrus cloud may have changed over the stretch of time over which we took the 4 exposures. We therefore did not consider the exposures in this epoch for absolute photometry. On the other hand, the other two epochs differed by as much as 0.02 and this may be an indication of intrinsic variability. Our lack of $I$ band detection imposes a limiting companion spectral type of M5V.

\subsection{IsWe 2}
\label{ssec:IsWe2}
We are reasonably confident that this star is variable at the 3-4 percent level, although the clear trend in the $B$, $V$ and $R$ bands is not clearly evident in the $I$ band (Table~\ref{tab:individualmagnitudes}). However, we noted that the $I$ band photometry taken in the second epoch was 30 minutes later than the other filters, which may have contributed to the lack of a similar trend. From $I$ band photometry we impose a limit of M4V on a companion's spectral type. The error on the average of 5 epochs is possibly larger than it should be because of the variability.

\subsection{K 1-13}

This is a faint elliptical/bipolar nebula also known as Abell~25.  The \citet{Schlafly2011} asymptotic reddening in this direction is $E(B-V)$ = 0.03, so the central star is essentially unreddened.  We adopt an integrated H$\alpha$ flux of $\log F = -11.95$ from \citet{Frew2012}, and \citet{Kaler1990} gave  an upper limit for the He~{\sc ii} $F$($\lambda$4686) flux of 0.5 $\times$ $F$(H$\beta$).  Using $V = 18.42$ for the central star, we obtain a hydrogen Zanstra temperature of 49\,kK, which is a lower limit.  The He~{\sc ii} flux limit can be used to derive an upper limit on the temperature of 92\,kK for this optically thick PN.  We have therefore adopted a temperature of 80$\pm$30\,kK.  We note that \citet{Abell1966} gave $V = 18.94$, $B-V = -0.14$, $U-B = -1.15$, so his $V$-band magnitude is quite different to ours. We impose a limiting spectral type of M3V from $I$ band photometry.

\subsection{K 2-2}
The ionising star has a rather low mass \citep{Napiwotzki1999} and the inferred post-AGB lifetime is not consistent with the kinematic age of the nebula.  This object may turn out to be another case of ionised ISM.  Like EGB~1, the nebula has a very low line width and a systemic nebular velocity close to zero LSR \citep{Lopez2012}.  Alternatively, this PN may be a senile example of a common-envelope ejection event.  \citet{Afsar2005} have suggested that the central star is a likely close binary, based on radial velocity variability. Our $J$ band photometry imposes a limiting spectral type for a companion of M4V.

\subsection{NGC 3587}
This nebula is also known as M~97. The $I$-band magnitude of \citet{Ciardullo1999} is fainter than ours by about one magnitude. We suspect it to be erroneous based on the fact that all available photometry ({\it Sloan Digital Sky Survey, 2MASS and Spitzer/IRAC}) is in line with our $B$, $V$ and $I$ values. We derive a companion spectral type limit of M6V from $J$ band photometry. 

\subsection{NGC 6853}
This beautiful, bright  PN, also known as M~27, is at a distance  of 405 $^{+28}_{-25}$ pc \citep{Benedict2009}.  \citet{Ciardullo1999}, discuss the companion to the central star discovered by \citet{Cudworth1973}. Assuming a physical association, a tentative spectroscopic parallax of 430 pc is derived, in agreement with the trigonometric distance, but a radial velocity measurement for the companion is needed. We detect an M5V companion with $J$ band photometry at the 1.5-$\sigma$ level. 

\subsection{PuWe 1}
\label{ssec:PuWe1}
The effective temperature of the central star of PuWe 1 was a weighted average of the determination by Napiwotzki (1999), Good et al. (2004) and Gianninas et al. (2010). Harris et al. (2007) give a trigonometric distance of 365$^{+47}_{-37}$ pc. \citet{Ciardullo1999} discuss the nature of a pair of stars (likely a binary) that are located at 5.2 arcsec from PuWe~1's central star.  This pair may be a wide companion to the central star, in which case this system would be a triple. However, the proper motion of the pair appears different to that of the central star, which would make the central star not associated with it. Our photometry excludes the optical pair. Our $I$ band imposes a limit for a companion spectral type of later than M4V.

\subsection{Sh 2-78}
\label{ssec:Sh2-78}
This star exhibits variability at the 0.3-mag level over the course of the 8 nights. The photometric magnitudes reveal only a $\sim$0.05-mag brightening between nights 4 and 5. We detected an M4V companion in the $I$ band at the 1.7-$\sigma$ level.

\subsection{Sh 2-176}
\label{ssec:Sh2-176}
This object was detected to be variable at the 5\% level from our monitoring dataset. Our three epochs of photometric observations reveal that indeed the object brightened between the first two epochs and dimmed thereafter. Although the data in all four filters tell a consistent story, the third epoch data are affected by a larger uncertainty and the $B$ band magnitude is quite a bit dimmer than the other filters. On the grounds of variability and larger errors we excluded the third epoch from the calculation of the average magnitudes (Table~\ref{tab:results}). By doing so the $I$ band excess is increased and the error is greatly decreased,  resulting in a 1.2-$\sigma$ detection of an M5V companion (see the two entries in Table~\ref{tab:resultsI}).

\subsection{Ton 320}
\label{ssec:Ton320}
The effective temperature of this central star of Ton~320 was determined as the weighted mean of the estimates of Good et al. (2004), \citet{Tremblay2011} and Gianninas et al. (2010), where the gravity was obtained from Gianninas et al. (2010).
The ionising star of this very faint PN (also known as TK~1) has a 2MASS K-band excess \citep{Holberg2005} but this may be spurious since the error bars on that value are large. The most stringent limit on the companion spectral type is imposed by our $J$ band observations (M6V).

\subsection{WeDe 1}
There is a faint wide companion to the central star of the PN WeDe~1 (also known as WDHS~1; \citealt{Weinberger1983}), though it is likely to be an optical companion. Our $I$ band photometry cannot detect a companion fainter than spectral type M5V.

\section{Conclusions and discussion}
\label{sec:conclusions}

In this work we have started a survey of the 2-kpc, volume limited sample to determine the binary fraction via a technique able to detect $I$ and $J$ band flux excess due to the presence of a cool companion. For this first survey we have selected 30 central stars of PN, of which we later determined three to be PN mimics. Of the remaining 27 central stars we have detected 8 with a possible $I$-band excesses (of which one was a previously known binary). For 12 of these objects we collected the best $J$ band photometry from the literature and determined that 6 out of 11 bona fide central stars have a flux excess. In total we have detected an excess in 11 of the 27 central stars in either of the two bands. 

For the $I$ band survey we calculate a detected fraction of 30\%, which when de-biased to account for undetected objects, to result in an unbiased binary fraction of 67-78\%. There are three principal de-biasing factors. The first to account for the fact that we do not detect companions fainter than M3-4V. The second, to account for the fact that we do not detect wide binaries by design. The third, to account for the fact that main sequence binaries that go through a common envelope on the RGB do not ascend the AGB and do not become PN. These factors have to be included in order to compare the PN binary fraction with the main sequence one.

From the $J$ band survey we calculate a binary fraction of  54\% which, when debiased to account for binaries at any period and all companions, becomes 100-107\%, clearly showing that the error bar due to the low number statistics is high. The discrepancy between the $I$ and $J$ band fractions is likely an effect of the low number statistics. Since the binary fraction determined using $I$ band photometry relies on more data than using the $J$ band photometry, it is likely that the $I$ band fraction is more accurate (even if the $I$ band is less sensitive). However the $J$ band fraction agrees with the preliminary work of \citet{Frew2007}, who used 34 objects (although their detection limits were poorly quantified). 

Thus debiased we can compare these fractions with that of F6V-G2V main sequence star binaries of (50$\pm$4)\% \citep{Raghavan2010}. We preliminarily conclude that there may be an overabundance of central star binaries, compared to the putative parent population.

We have also noticed how, starting from the main sequence binary fraction and period distribution we would expect only a few percent of central stars of PN to be post-common envelope binaries, whereas previous surveys detected 15-20\%. This discrepancy can only be reconciled within the {\it binary hypothesis}.  

Here we have demonstrated that the binary fraction is an elusive quantity. Even the best, most accurate observations cannot easily detect faint companions and the only way to reduce the error is by analysing a much larger sample with similarly accurate photometry. Such high accuracy observations are extremely hard to obtain because of the need of photometric weather and the importance to obtain several observations on different nights. In the second instalment of this paper we will use the same technique with an additional dataset and combine the results. Our goal is to analyse with similar accuracy the entire 2.5-kpc volume-limited sample of Frew (2008), comprising approximately 250 central stars.


Finally, we compare the binary fraction determined here with two alternative predictions. The first one was done on the basis of PN morphology alone. \citet{DeMarco2011b}, revising the scheme of \citet{Soker1997}, predicted that about 60\% of all central stars have interacted with a stellar companion. Here we measured that 48\% of central stars have a companion closer than 1300~AU (30\% $\times$ 1.45-1.69, where we took the middle of the range; \S~\ref{sssec:completion}). These two numbers are easily the same within the uncertainty, but only if all of the companions we detected (and those we did not detect but accounted for) are closer than $\sim$100~AU, i.e., have interacted with the central star. The comparison is possibly closer using the $J$ band sample, whereby 67\% of all central stars are binaries with any separation to a distance of $\sim$1000~AU. 

Another prediction was that of \citet{Moe2011}, who used a population synthesis analysis to predict that a fraction of 70\% of all central star derives from a binary interaction. This fraction is even higher and more discrepant with the fraction detected here for the $I$ band sample, but more in line with that derived from the $J$ band sample.  All these comparisons will need to be carried out again, once the PN binary fraction is finalised and a period distribution is obtained.

\section*{Acknowledgments} OD and J-CP acknowledge funding from NSF grant AST-0607111 which was used during the initial phases of this work. J-CP is thankful of Falk Herwig's support. We thank Thomas Rauch for his help in obtaining the best synthetic stellar atmospheres. We thank Dimitri Douchin for expediting his analysis of the variability properties of the stars in our sample. We are grateful to Howard Bond for helping in the observing proposal writing stage. The TheoSSA service (http:\/\/dc.g-vo.org\/theossa) used to retrieve theoretical spectra for this paper was constructed as part of the activities of the German Astrophysical Virtual Observatory.

\bibliography{convert,bibliography}

\appendix
\section{Individual photometric magnitudes of the central star of PN sample}
\label{app:individual magnitudes}

In Table~\ref{tab:individualmagnitudes} we report all the photometric measurements for our central stars of PN. The measurements for each star were averaged according to the Equations \ref{eq:mean} and \ref{eq:meanerror} in \S~\ref{sec:measurements} and presented in Table~\ref{tab:results}. 

\begin{table*}
\scalebox{0.85}
{\begin{tabular}{lllllllllll}
\hline
Name&Night&B&V&R&I&$\sigma_B$&$\sigma_V$&$\sigma_R$&$\sigma_I$&Airmass\\
\hline
     A~7  &    1  &    15.195  &    15.496  &    15.620  &     --  &     0.014  &     0.012  &     0.011  &     --  &       1.6 \\
         &    5  &    15.189  &    15.500  &    15.652  &    15.830  &     0.013  &     0.010  &     0.020  &     0.025  &       1.6 \\
         &    7  &    15.188  &    15.491  &    15.635  &    15.808  &     0.012  &     0.009  &     0.019  &     0.021  &       1.5 \\
    A~16  &    4  &    18.513  &    18.718  &    18.725  &    18.671  &     0.013  &     0.017  &     0.020  &     0.023  &       1.2 \\
         &    5  &    18.535  &    18.729  &    18.758  &    18.707  &     0.015  &     0.013  &     0.021  &     0.028  &       1.4 \\
         &    6  &    18.497  &    18.684  &    18.710  &    18.679  &     0.022  &     0.023  &     0.027  &     0.049  &       1.1 \\
    A~20  &    4  &    16.214  &    16.468  &    16.551  &    16.681  &     0.011  &     0.014  &     0.018  &     0.018  &       1.2 \\
         &    5  &    16.218  &    16.472  &    16.568  &    16.716  &     0.015  &     0.010  &     0.020  &     0.025  &       1.5 \\
         &    7  &    16.216  &    16.459  &    16.559  &    16.710  &     0.012  &     0.010  &     0.019  &     0.020  &       1.5 \\
    A~28  &    5  &    16.281  &    16.565  &    16.700  &    16.880  &     0.013  &     0.010  &     0.019  &     0.024  &       1.3 \\
         &    6  &    16.290  &    16.555  &    16.682  &    16.891  &     0.017  &     0.012  &     0.018  &     0.037  &       1.2 \\
         &    7  &    16.268  &    16.541  &    16.693  &    16.854  &     0.020  &     0.022  &     0.025  &     0.046  &       1.1 \\
    A~31  &    5  &    15.195  &    15.545  &    15.702  &    15.849  &     0.014  &     0.010  &     0.020  &     0.024  &       1.2 \\
         &    7  &    15.207  &    15.543  &    15.684  &    15.816  &     0.012  &     0.009  &     0.018  &     0.021  &       1.4 \\
    A~57  &    4  &    17.904  &    17.747  &    17.460  &    17.212  &     0.014  &     0.018  &     0.019  &     0.021  &       1.3 \\
         &    7  &    17.902  &    17.724  &    17.440  &    17.208  &     0.014  &     0.015  &     0.022  &     0.023  &       1.6 \\
    A~71  &    5  &    19.380  &    19.339  &    19.261  &    19.107  &     0.014  &     0.015  &     0.025  &     0.035  &       1.1 \\
         &    6  &    19.410  &    19.324  &    19.237  &    19.274  &     0.022  &     0.025  &     0.032  &     0.067  &       1.3 \\
         &    7  &    19.372  &    19.338  &    19.258  &    19.201  &     0.013  &     0.013  &     0.024  &     0.038  &       1.1 \\
  A~72        &    3  &    15.769  &    16.095  &    16.398  &    16.472  &     0.021  &     0.022  &     0.032  &     0.053  &       1.2 \\
         &    4  &    15.740  &    16.028  &    16.165  &    16.358  &     0.013  &     0.016  &     0.020  &     0.024  &       1.6 \\
         &    7  &    15.776  &    16.085  &    16.211  &    16.364  &     0.013  &     0.010  &     0.019  &     0.023  &       1.5 \\
    A~79  &    3  &    17.813  &    16.961  &    16.404  &    15.814  &     0.015  &     0.014  &     0.018  &     0.026  &       1.2 \\
         &    4  &    17.838  &    16.972  &    16.417  &    15.607  &     0.021  &     0.023  &     0.033  &     0.053  &       1.1 \\
         &    6  &    17.829  &    16.964  &    16.340  &    15.733  &     0.022  &     0.026  &     0.060  &     0.055  &       1.4 \\
      A~84    &    3  &    18.386  &    18.604  &    18.640  &    18.669  &     0.023  &     0.024  &     0.033  &     0.055  &       1.1 \\
         &    4  &    18.357  &    18.590  &    18.621  &    18.650  &     0.012  &     0.016  &     0.020  &     0.024  &       1.1 \\
         &    7  &    18.364  &    18.571  &    18.590  &    18.693  &     0.011  &     0.011  &     0.019  &     0.023  &       1.1 \\
   EGB~1  &    4  &    16.313  &    16.445  &    16.446  &    16.463  &     0.012  &     0.014  &     0.019  &     0.019  &       1.4 \\
         &    5  &    16.301  &    16.434  &    16.460  &    16.507  &     0.015  &     0.011  &     0.022  &     0.024  &       1.4 \\
   EGB~6  &    4  &    15.691  &    16.001  &    16.129  &    16.292  &     0.011  &     0.014  &     0.018  &     0.018  &       1.2 \\
         &    7  &    15.694  &    15.997  &    16.145  &    16.311  &     0.012  &     0.009  &     0.018  &     0.023  &       1.5 \\
  HaWe~5  &    4  &    17.310  &    17.444  &    17.469  &    17.536  &     0.012  &     0.011  &     0.010  &     0.015  &       1.0 \\
         &    5  &    17.333  &    17.434  &    17.477  &    17.513  &     0.012  &     0.012  &     0.025  &     0.026  &       1.2 \\
   HDW~3  &    1  &    17.081  &    17.188  &    17.194  &    17.218  &     0.012  &     0.010  &     0.009  &     0.010  &       1.0 \\
         &    4  &    17.086  &    17.192  &    17.287  &    17.249  &     0.011  &     0.013  &     0.018  &     0.023  &       1.0 \\
         &    7  &    17.085  &    17.183  &    17.200  &    17.238  &     0.010  &     0.009  &     0.016  &     0.022  &       1.1 \\
  IsWe~1  &    5  &    16.392  &    16.531  &    16.586  &    16.659  &     0.013  &     0.011  &     0.018  &     0.024  &       1.2 \\
         &    6  &    16.367  &    16.517  &    16.556  &    16.630  &     0.012  &     0.011  &     0.016  &     0.018  &       1.1 \\
         &    7  &    16.360  &    16.517  &    16.592  &    16.650  &     0.020  &     0.022  &     0.025  &     0.046  &       1.1 \\
 IsWe~2        &    3  &    18.129  &    18.173  &    18.114  &    18.052  &     0.021  &     0.023  &     0.034  &     0.055  &       1.2 \\
         &    4  &    18.124  &    18.168  &    18.120  &    18.101  &     0.024  &     0.015  &     0.020  &     0.025  &       1.2 \\
         &    5  &    18.177  &    18.200  &    18.160  &    18.092  &     0.013  &     0.012  &     0.020  &     0.028  &       1.2 \\
         &    6  &    18.157  &    18.156  &    18.101  &    18.139  &     0.022  &     0.023  &     0.029  &     0.050  &       1.4 \\
         &    7  &    18.115  &    18.113  &    18.088  &    18.102  &     0.013  &     0.011  &     0.020  &     0.025  &       1.3 \\
  JnEr~1  &    4  &    16.770  &    17.143  &    17.288  &    17.483  &     0.018  &     0.014  &     0.018  &     0.020  &       1.1 \\
         &    7  &    16.778  &    17.117  &    17.287  &    17.531  &     0.014  &     0.010  &     0.018  &     0.033  &       1.2 \\
  K~1-13  &    3  &    18.062  &    18.425  &    18.569  &    18.798  &     0.013  &     0.017  &     0.021  &     0.027  &       1.3 \\
         &    4  &    18.045  &    18.420  &    18.603  &    18.896  &     0.013  &     0.013  &     0.020  &     0.030  &       1.2 \\
         &    5  &    18.036  &    18.435  &    18.613  &    18.852  &     0.033  &     0.026  &     0.037  &     0.061  &       1.2 \\
   K~2-2  &    4  &    13.984  &    14.278  &    14.396  &    14.536  &     0.011  &     0.013  &     0.017  &     0.020  &       1.1 \\
         &    5  &    13.974  &    14.261  &    14.395  &    14.561  &     0.012  &     0.010  &     0.020  &     0.025  &       1.4 \\
         &    7  &    13.972  &    14.254  &    14.379  &    14.564  &     0.011  &     0.010  &     0.018  &     0.019  &       1.3 \\
NGC~3587  &    4  &    15.414  &    15.771  &    15.953  &    16.164  &     0.013  &     0.014  &     0.019  &     0.020  &       1.3 \\
         &    5  &    15.415  &    15.788  &    15.963  &    16.227  &     0.014  &     0.012  &     0.018  &     0.027  &       1.3 \\
         &    6  &    15.413  &    15.768  &    15.966  &    16.207  &     0.021  &     0.023  &     0.027  &     0.047  &       1.3 \\
NGC~6720  &    4  &    15.393  &    15.793  &    15.898  &    16.050  &     0.012  &     0.020  &     0.020  &     0.019  &       1.2 \\
         &    7  &    15.420  &    15.747  &    15.904  &    16.075  &     0.016  &     0.019  &     0.022  &     0.022  &       1.6 \\
NGC~6853  &    4  &    13.725  &    14.075  &    14.244  &    14.404  &     0.012  &     0.016  &     0.019  &     0.020  &       1.3 \\
         &    5  &    13.777  &    14.100  &    14.241  &    14.390  &     0.012  &     0.013  &     0.019  &     0.045  &       1.1 \\
         &    7  &    13.744  &    14.090  &    14.255  &    14.417  &     0.013  &     0.013  &     0.020  &     0.035  &       1.6 \\
  PuWe~1  &    5  &    15.298  &    15.551  &    15.673  &    15.802  &     0.013  &     0.010  &     0.019  &     0.023  &       1.4 \\
         &    7  &    15.285  &    15.539  &    15.651  &    15.785  &     0.011  &     0.010  &     0.018  &     0.018  &       1.3 \\
 Sh~2-78  &    4  &    17.643  &    17.666  &    17.630  &    17.567  &     0.014  &     0.016  &     0.030  &     0.038  &       1.4 \\
         &    5  &    17.624  &    17.655  &    17.580  &    17.513  &     0.013  &     0.012  &     0.038  &     0.046  &       1.3 \\
 Sh~2-176        &    3  &    18.639  &    18.580  &    18.597  &    18.602  &     0.022  &     0.024  &     0.036  &     0.059  &       1.1 \\
  &    4  &    18.460  &    18.569  &    18.571  &    18.521  &     0.018  &     0.017  &     0.020  &     0.029  &       1.1 \\
         &    7  &    18.431  &    18.536  &    18.553  &    18.541  &     0.011  &     0.015  &     0.020  &     0.028  &       1.1 \\
Sh~2-188  &    4  &    17.438  &    17.454  &    17.407  &    17.361  &     0.012  &     0.014  &     0.017  &     0.019  &       1.1 \\
         &    5  &    17.416  &    17.446  &    17.402  &    17.381  &     0.012  &     0.009  &     0.018  &     0.025  &       1.2 \\
        &    7  &    17.419  &    17.444  &    17.385  &    17.386  &     0.013  &     0.009  &     0.017  &     0.017  &       1.1 \\
 Ton~320  &    5  &    15.385  &    15.731  &    15.900  &    16.126  &     0.013  &     0.011  &     0.020  &     0.025  &       1.3 \\
         &    7  &    15.373  &    15.718  &    15.880  &    16.091  &     0.015  &     0.012  &     0.019  &     0.037  &       1.2 \\
  WeDe~1  &    5  &    16.961  &    17.229  &    17.329  &    17.503  &     0.018  &     0.011  &     0.023  &     0.028  &       1.7 \\
         &    6  &    16.962  &    17.221  &    17.334  &    17.470  &     0.012  &     0.011  &     0.018  &     0.027  &       1.3 \\      
         &    7  &     16.946  & 17.231 & 17.354 & 17.500 & 0.021 & 0.022 & 0.027 & 0.046 & 1.1\\
\hline
\end{tabular}}
\caption{The individual photometric magnitudes for our sample of PN central stars (including three PN mimics). The date and time corresponding to each night can be found in Table~\ref{tab:logs}. DeHt~5 and HDW~4, having only one epoch of data are reported only in Table~\ref{tab:results} \label{tab:individualmagnitudes} }
\end{table*}

\section{The theoretical Johnson-Cousins colours of hot stars}
\label{app:stellarmodels}

In this appendix we report the colours determined for a grid of stellar atmosphere models with Solar and PG1159 abundances. The theoretical stellar atmosphere models were calculated with the simulation code TMAW, the web interface to TMAP \citep{Werner1999,Werner2003,Rauch2003}, or the German Astrophysical Virtual Observatory grid calculations TheoSSA\footnote{dc.zah.uni-heidelberg.de/theossa/} which are carried out for standard, non-tailored abundances but typically with a more complete model atom. TheoSSA is deemed more correct but differences were typically lower than 1\%. The TMAW stellar atmosphere models are calculated for a solar composition (with mass fractions $\beta_H$=0.71, $\beta_{He}$=0.28 $\beta_C$=0.001 $\beta_N$=0.002 $\beta_O$=0.006). The TheoSSA models are for $\beta_H$=0.7, $\beta_{He}$=0.3 unless indicated {\it or} for the composition typical of hydrogen-deficient PG1159 stars (\citealt{Werner2006}; $\beta_{He}$=0.33 $\beta_C$=0.50 $\beta_N$=0.02 $\beta_O$=0.15). The colours are then determined using the routine {\it calcphot} within the {\it iraf} package (using the {\it vegamag} and {\it effstim} options). The filter system used to obtain the synthetic colours is the Johnson-Cousins system, the same used for the determination of the standard magnitudes by \citet{Landolt1992}. These colours were modified by the zero points listed in table 3.1 of the {\it synphot} manual, 0.010 for the $B-V$ and $R-I$ colours and $-0.002$ for the $V-I$ one \citep{MaizApellaniz2004,Holberg2006}. The $V-J$ and $J-H$ colours have not been corrected. We note that the offset for the $V$ band alone is 0.026 mag. If the correction for the $J$ magnitude were zero, this would imply that the theoretical colours are redder, and the resulting excess smaller by about 3\%. We note that none of our detections in the $V-J$ band would be affected, although the detected spectral types would be slightly fainter. The theoretical colours for Solar as well as PG1159 abundances are tabulated in Table~\ref{tab:theoreticalcolours} as a function of temperature and gravity, where we have also tabulated blackbody values for comparison. The parameter space covered is that which applies to central star of PN as well as post-RGB, horizontal branch stars. See for instance the $\log g - T_{\rm eff}$ diagram in \citet{Napiwotzki1999}. 

\begin{table*}
\scalebox{0.80}
{\begin{tabular}{llllllll}
\hline
$T_{eff}$&$\log g$&$B-V$&$V-I$&$V-J$&$R-I$&$J-H$&Abundance\\
(kK)&&(mag)&(mag)&(mag)&(mag)&(mag)\\
\hline
20	&	4	&	-0.195	&	-0.193	&	-0.466	&	-0.097	&	-0.073	&	Solar (TMAW)\\
20	&	5	&	-0.180	&	-0.199	&	-0.475	&	-0.099	&	-0.076	&	Solar (TMAW)\\
20	&	--   	&	-0.139	&	-0.118	&	-0.414	&	-0.066	&	-0.068	&	BB\\
30	&	4	&	-0.269	&	-0.277	&	-0.653	&	-0.145	&	-0.109	&	Solar (TMAW)\\
30	&	5	&	-0.260	&	-0.280	&	-0.660	&	-0.145	&	-0.111	&	Solar (TMAW)\\
30	&	7	&	-0.221	&	-0.289	&	-0.680	&	-0.143	&	-0.106	&	Solar (TMAW)\\
30	&	8	&	-0.191	&	-0.289	&	-0.695	&	-0.139	&	-0.100	&	Solar (TMAW)\\
30	&	--   	&	-0.226	&	-0.214	&	-0.582	&	-0.116	&	-0.098	&	BB\\
40	&	4	&	-0.271	&	-0.281	&	-0.684	&	-0.150	&	-0.114	&	H0.728He0.249\\
40	&	5	&	-0.279	&	-0.298	&	-0.713	&	-0.158	&	-0.119	&	H0.894He0.106\\
40	&	6	&	-0.274	&	-0.304	&	-0.724	&	-0.161	&	-0.121	&	H0.738He0.014\\
40	&	7	&	-0.277	&	-0.319	&	-0.759	&	-0.165	&	-0.126	&	Solar (TMAW)\\
40	&	8	&	-0.263	&	-0.318	&	-0.759	&	-0.161	&	-0.121	&	Solar (TMAW)\\
40	&	--  	&	-0.266	&	-0.259	&	-0.662	&	-0.140	&	-0.113	&	BB\\
50	&	4	&	-0.280	&	-0.291	&	-0.704	&	-0.156	&	-0.116	&	H0.738He0.249\\
50	&	5	&	-0.290	&	-0.305	&	-0.725	&	-0.162	&	-0.119	&	H1.0\\
50	&	6	&	-0.298	&	-0.310	&	-0.745	&	-0.164	&	-0.125	&	Solar (TMAW)\\
50	&	7	&	-0.294	&	-0.316	&	-0.758	&	-0.166	&	-0.127	&	Solar (TMAW)\\
50	&	8	&	-0.280	&	-0.315	&	-0.760	&	-0.166	&	-0.126	&	H0.7He0.3\\
50	&	--  	&	-0.289	&	-0.285	&	-0.708	&	-0.154	&	-0.122	&	BB\\
60	&	5	&	-0.302	&	-0.315	&	0.753	&	-0.169	&	-0.126	&	H0.7He0.3\\
60	&	6	&	-0.307	&	-0.319	&	-0.755	&	-0.169	&	-0.129	&	Solar (TMAW)\\
60	&	7	&	-0.305	&	-0.322	&	-0.770	&	-0.170	&	-0.130	&	Solar (TMAW)\\
60	&	8	&	-0.295	&	-0.322	&	-0.771	&	-0.170	&	-0.129	&	H0.7He0.3\\
60	&	--  	&	-0.304	&	-0.302	&	-0.737	&	-0.163	&	-0.128	&	BB\\
70	&	5	&	-0.310	&	-0.325	&	-0.771	&	-0.174	&	-0.130	&	H0.7He0.3\\
70	&	6	&	-0.312	&	-0.323	&	-0.722	&	-0.171	&	-0.131	&	Solar (TMAW)\\
70	&	7	&	-0.311	&	-0.328	&	-0.780	&	-0.173	&	-0.132	&	Solar (TMAW)\\
70	&	8	&	-0.306	&	-0.326	&	-0.783	&	-0.175	&	-0.132	&	H0.6He0.4\\
70	&	--   	&	-0.314	&	-0.314	&	-0.760	&	-0.169	&	-0.132	&	BB\\
80	&	5	&	-0.316	&	-0.331	&	-0.781	&	-0.177	&	-0.130	&	H0.7He0.3\\
80	&	6	&	-0.318	&	-0.332	&	-0.787	&	-0.178	&	-0.133	&	H0.7He0.3\\
80	&	7	&	-0.318	&	-0.333	&	-0.791	&	-0.176	&	-0.135	&	Solar (TMAW)\\
80	&	8	&	-0.314	&	-0.334	&	-0.797	&	-0.179	&	-0.135	&	H0.7He0.3\\
80	&	--  	&	-0.322	&	-0.323	&	-0.774	&	-0.174	&	-0.135	&	BB\\
90	&	5	&	-0.316	&	-0.330	&	-0.779	&	-0.177	&	-0.129	&	H0.7He0.3\\
90	&	6	&	-0.326	&	-0.340	&	-0.806	&	-0.181	&	-0.140	&	Solar (TMAW)\\
90	&	7	&	-0.325	&	-0.340	&	-0.804	&	-0.180	&	-0.138	&	Solar (TMAW)\\
90	&	8	&	-0.322	&	-0.341	&	-0.808	&	-0.183	&	-0.137	&	H0.7He0.3\\
90	&	-- 	&	-0.328	&	-0.329	&	-0.786	&	-0.177	&	-0.137	&	BB\\
100	&	6	&	-0.332	&	-0.347	&	-0.817	&	-0.184	&	-0.141	&	Solar (TMAW)\\
100	&	7	&	-0.331	&	-0.346	&	-0.814	&	-0.183	&	-0.139	&	Solar (TMAW)\\
100	&	8	&	-0.329	&	-0.345	&	-0.814	&	-0.185	&	-0.138	&	H0.7He0.3\\
100	&	 --	&	-0.332	&	-0.335	&	-0.796	&	-0.180	&	-0.139	&	BB\\
110	&	6	&	-0.330	&	-0.341	&	-0.801	&	-0.182	&	-0.132	&	H0.7He0.3\\
110	&	7	&	-0.336	&	-0.349	&	-0.819	&	-0.185	&	-0.140	&	Solar (TMAW)\\
110	&	8	&	-0.334	&	-0.347	&	-0.815	&	-0.186	&	-0.138	&	H0.7He0.3\\
110	&	-- 	&	-0.336	&	-0.339	&	-0.803	&	-0.182	&	-0.140	&	BB\\
120	&	6	&	-0.332	&	-0.343	&	-0.803	&	-0.183	&	-0.132	&	H0.7He0.3\\
120	&	7	&	-0.339	&	-0.352	&	-0.822	&	-0.186	&	-0.141	&	Solar (TMAW)\\
120	&	7.5	&	-0.337	&	-0.347	&	-0.812	&	-0.186	&	-0.136	&	H0.7He0.3\\
120	&	--	&	-0.339	&	-0.343	&	-0.810	&	-0.184	&	-0.142	&	BB\\
130	&	6	&	-0.333	&	-0.344	&	-0.804	&	-0.183	&	-0.132	&	H0.7He0.3\\
130	&	7	&	-0.341	&	-0.353	&	-0.827	&	-0.187	&	-0.142	&	Solar (TMAW)\\
130	&	8	&	-0.340	&	-0.350	&	-0.818	&	-0.187	&	-0.137	&	H0.7He0.3\\
130	&	--	&	-0.342	&	-0.346	&	-0.815	&	-0.186	&	-0.143	&	BB\\
140	&	6	&	-0.335	&	-0.345	&	-0.805	&	-0.184	&	-0.132	&	H0.7He0.3\\
140	&	7	&	-0.342	&	-0.354	&	-0.830	&	-0.188	&	-0.143	&	Solar (TMAW)\\
140	&	8	&	-0.343	&	-0.351	&	-0.819	&	-0.188	&	-0.137	&	H0.6He0.4\\
140	&	--	&	-0.344	&	-0.348	&	-0.820	&	-0.187	&	-0.144	&	BB\\
150	&	6	&	-0.335	&	-0.346	&	-0.808	&	-0.184	&	-0.132	&	H0.7He0.3\\
150	&	7	&	-0.344	&	-0.355	&	-0.831	&	-0.188	&	-0.143	&	Solar (TMAW)\\
150	&	8	&	-0.343	&	-0.352	&	-0.820	&	-0.188	&	-0.137	&	H0.8He0.2\\
160	&	6	&	-0.336	&	-0.347	&	-0.810	&	-0.185	&	-0.132	&	H0.7He0.3\\
160	&	7	&	-0.345	&	-0.356	&	-0.833	&	-0.189	&	-0.144	&	Solar (TMAW)\\
160	&	8	&	-0.345	&	-0.353	&	-0.820	&	-0.188	&	-0.136	&	H0.7He0.3\\
160	&	6	&	-0.335	&	-0.349	&	-0.814	&	-0.185	&	-0.134	&	H0.8He0.2\\
170	&	7	&	-0.346	&	-0.358	&	-0.837	&	-0.190	&	-0.145	&	Solar (TMAW)\\
170	&	8	&	-0.346	&	-0.353	&	-0.821	&	-0.189	&	-0.136	&	H0.7He0.3\\
170	&	--	&	-0.349	&	-0.354	&	-0.830	&	-0.190	&	-0.146	&	BB\\
\hline
100	&	6	&	-0.307	&	-0.321	&	-0.769	&	-0.178	&	-0.151	&	He0.33C0.50N0.02O0.15\\
100	&	7	&	-0.340	&	-0.357	&	-0.835	&	-0.194	&	-0.151	&	He0.33C0.50N0.02O0.15\\
110	&	6	&	-0.369	&	-0.351$^1$    &	-0.987	&	-0.231	&	-0.204	&	He0.33C0.50N0.02O0.15\\
110	&	7	&	-0.346	&	-0.355	&	-0.832	&	-0.194	&	-0.148	&	He0.33C0.50N0.02O0.15\\
120	&	6	&	-0.386	&	-0.382	&	-0.876	&	-0.203	&	-0.143	&	He0.33C0.50N0.02O0.15\\
120	&	7	&	-0.350	&	-0.357	&	-0.836	&	-0.195	&	-0.148	&	He0.33C0.50N0.02O0.15\\
130	&	6	&	-0.387	&	-0.384	&	-0.878	&	-0.204	&	-0.145	&	He0.33C0.50N0.02O0.15\\
130	&	7	&	-0.352	&	-0.359	&	-0.838	&	-0.196	&	-0.147	&	He0.33C0.50N0.02O0.15\\
\hline
\multicolumn{8}{l}{$^1$This colour was interpolated between the values for the 110kK and 130kK atmospheres.}
\end{tabular}}
\caption{Predicted colours of single post-AGB stars using TMAP models (TMAW with solar metallicity - see text, or TheoSSA, where the model atom and abundances are indicated as mass fractions adjacent to the relevant atom) as well as for blackbody curves (BB). These values have been adjusted for the small corrections of 0.010~mag for the $B-V$ and $R-I$ colours and --0.002~mag for the $V-I$ colours \label{tab:theoreticalcolours}}
\end{table*}

\section{The magnitudes, colours and masses of main sequence stars}
\label{app:coolstars}

In this Appendix we report the value for absolute $V$ magnitudes and colours for main sequence stars of spectral type B3V to L0V. These data were used to interpret the $I$ and $J$ band excesses we detected in terms of companion spectral type. 

For the colours of the companions (assumed to be main sequence stars) we have used a variety of data from the literature.  For spectral types earlier than B8V, we averaged the $V$-band absolute magnitudes and $B-V$ colours  from \citet{Schmidt-Kaler1982} and \citet{Wegner1994}, with the data from Eric Mamajek's website\footnote{http://www.pas.rochester.edu/\mytilde emamajek/EEM$\textunderscore$dwarf$\textunderscore$UBVIJHK$\textunderscore$colors$\textunderscore$Teff.dat} (last updated in Dec 2011).   For spectral types in the range B8V--K7V, we averaged the data from Mamajek's website, \citet{Schmidt-Kaler1982} and \citet{Kraus2007}, transformed following \citet{Lupton2005}\footnote{http://www.sdss.org/dr5/algorithms/sdssUBVRITransform.html}, with additional $B-V$ data from \citet{Bessel1990} and \citet{Bessell1991}. We averaged the $M_V$ and $B-V$ data from \citet{Leggett1992}, \citet{Kirkpatrick1994}, Mamajek's website and Frew (unpublished) for spectral types later than M0V.

The $V-R_{c}$ and colours were taken from \citet{Bessell1979} or \citet{Bessel1990} for spectral types earlier than M0V, and from \citet{Leggett1992}, \citet{Kirkpatrick1994} and Frew (unpublished) for spectral types later than M0V; The $V-I_{c}$ colours were similarly averaged from \citet{Bessell1979}, \citet{Bessel1990}, \citet{Bessell1988}, \citet{Leggett1992}, \citet{Kirkpatrick1994}, Mamajek's website, and Frew (unpublished).  The $V-J$ and $V-H$ colours were derived from \citet{Kraus2007} and Mamajek's website for spectral types earlier than M0V, and from \citet{Kirkpatrick1994}, Mamajek's website and this work for spectral types later or equal to M0V.  Finally, the stellar masses were derived from the relations given in \citet{Henry1993} for spectral types later than A5V, together with other literature estimates for earlier spectral types.

\begin{table*}
{\footnotesize
\medskip
\begin{tabular}{cccccccccl}
\hline
Spec.    &   ~$M_V$       &      ~$(U-B)_0$~    &      $(B-V)_0$        &      $(V-R_{c})_0$     &     $(V-I_{c})_0$    &      $(V-J)_0$      &         $(V-H)_0$     &~$(V-K_s)_0$~ &   Mass      \\
type       &    (mag)          &          (mag)            &        (mag)             &        (mag)                 &        (mag)              &        (mag)          &        (mag)              &        (mag)        & M$_{\odot}$ \\
\hline
B2 V        &     $-2.10$     &      $-0.82$           &    $-0.23$        &    $-0.09$        &     $-0.23$        &    $-0.49$          &     $-0.57$        &    $-0.60$    &         7.7        \\
B3 V        &     $-1.45$     &      $-0.69$           &    $-0.19$        &    $-0.07$        &     $-0.19$        &    $-0.38$          &     $-0.47$        &    $-0.49$    &         5.7        \\
B4 V        &     $-1.22$     &      $-0.63$         &    $-0.18$        &    $-0.06$        &     $-0.18$        &    $-0.34$          &     $-0.43$        &    $-0.45$    &         5.1        \\
B5 V        &     $-1.06$     &      $-0.58$         &    $-0.16$        &    $-0.05$        &     $-0.17$        &    $-0.32$          &     $-0.41$        &    $-0.42$    &         4.8        \\
B6 V        &     $-0.77$     &      $-0.50$         &    $-0.15$        &    $-0.04$        &     $-0.15$        &    $-0.28$          &     $-0.35$        &    $-0.36$    &         4.2        \\
B7 V        &     $-0.53$     &       $-0.44$        &     $-0.13$        &    $-0.04$        &     $-0.11$        &    $-0.25$          &     $-0.32$        &    $-0.33$    &          4.0        \\
B8 V        &     $-0.24$     &       $-0.35$        &     $-0.10$        &    $-0.03$        &     $-0.09$        &    $-0.19$          &     $-0.26$        &    $-0.26$    &          3.5        \\
B9 V        &     $+0.16$     &       $-0.20$        &     $-0.07$        &    $-0.02$        &     $-0.06$        &    $-0.09$          &     $-0.14$        &    $-0.12$    &          3.1        \\
A0 V        &     $+0.79$     &       $-0.02$        &     $-0.01$        &    $+0.00$        &     $-0.01$        &    $+0.05$          &     $+0.01$        &    $+0.04$     &          2.4        \\
A1 V        &     $+1.05$     &       $+0.03$        &     $+0.03$        &    $+0.01$        &     $+0.03$        &    $+0.09$          &     $+0.07$        &    $+0.10$    &          2.2        \\
A2 V        &     $+1.36$     &       $+0.06$        &     $+0.06$        &    $+0.03$         &     $+0.07$        &    $+0.17$          &     $+0.16$        &    $+0.19$    &        2.1        \\
A3 V        &     $+1.53$     &       $+0.08$        &     $+0.09$        &    $+0.05$         &     $+0.10$        &    $+0.20$          &     $+0.20$        &    $+0.23$    &        2.0        \\
A4 V        &     $+1.74$     &       $+0.09$        &     $+0.13$        &    $+0.06$         &     $+0.17$        &    $+0.30$          &     $+0.32$        &    $+0.36$    &          1.95        \\
A5 V        &     $+1.90$     &       $+0.10$        &     $+0.15$        &    $+0.07$         &     $+0.19$        &    $+0.36$          &     $+0.37$        &    $+0.40$    &          1.89        \\
A7 V        &     $+2.16$     &       $+0.10$        &     $+0.20$        &    $+0.12$         &     $+0.24$        &    $+0.45$          &     $+0.49$        &    $+0.53$    &         1.76        \\
A8 V        &     $+2.35$     &       $+0.09$        &     $+0.25$        &    $+0.15$         &     $+0.29$        &    $+0.51$          &     $+0.59$        &    $+0.63$    &         1.65        \\
F0 V        &     $+2.63$     &       $+0.04$        &     $+0.30$        &    $+0.18$         &     $+0.36$        &    $+0.60$          &     $+0.69$        &    $+0.73$    &         1.58        \\
F2 V        &     $+3.00$     &       $-0.00$        &     $+0.36$        &    $+0.21$         &     $+0.43$        &    $+0.72$          &     $+0.88$        &    $+0.93$    &           1.46        \\
F5 V        &     $+3.46$     &       $-0.02$        &     $+0.43$        &    $+0.27$         &     $+0.52$        &    $+0.89$          &     $+1.03$        &    $+1.08$    &           1.31        \\
F8 V        &     $+4.01$     &       $+0.01$        &     $+0.53$        &    $+0.30$         &     $+0.60$        &    $+1.06$          &     $+1.23$        &    $+1.29$    &           1.19        \\
G0 V        &     $+4.40$     &       $+0.05$        &     $+0.58$        &    $+0.33$         &     $+0.67$        &    $+1.12$          &     $+1.36$        &    $+1.42$    &          1.10        \\
G2 V        &     $+4.72$     &       $+0.12$        &     $+0.64$        &    $+0.36$         &     $+0.71$        &    $+1.18$          &     $+1.47$        &    $+1.55$    &         1.03        \\
G5 V        &     $+5.07$     &       $+0.19$        &     $+0.67$        &    $+0.39$         &     $+0.74$        &    $+1.25$          &     $+1.56$        &    $+1.64$    &         0.99        \\
G8 V        &     $+5.51$     &       $+0.29$        &     $+0.74$        &    $+0.43$         &     $+0.78$        &    $+1.38$          &     $+1.69$        &    $+1.77$    &         0.93        \\
K0 V        &     $+5.89$     &       $+0.44$        &     $+0.82$        &    $+0.46$         &     $+0.85$        &    $+1.50$          &     $+1.86$        &    $+2.03$    &         0.87        \\
K1 V        &     $+6.08$     &       $+0.52$        &     $+0.85$        &    $+0.50$         &     $+0.88$        &    $+1.55$          &     $+1.94$        &    $+1.95$    &         0.85        \\
K2 V        &     $+6.37$     &       $+0.62$        &     $+0.90$        &    $+0.54$         &     $+0.93$        &    $+1.67$          &     $+2.06$        &    $+2.16$    &         0.82        \\
K3 V        &     $+6.61$     &       $+0.80$        &     $+0.98$        &    $+0.57$         &     $+1.05$        &    $+1.81$          &     $+2.30$        &    $+2.41$    &         0.80        \\
K5 V        &     $+7.34$     &       $+1.08$        &     $+1.16$        &    $+0.73$         &     $+1.29$        &    $+2.19$          &     $+2.71$        &    $+2.84$    &           0.72        \\
K7 V        &     $+8.16$     &       $+1.20$        &     $+1.34$        &    $+0.81$         &     $+1.57$        &    $+2.54$          &     $+3.24$        &    $+3.41$    &           0.65        \\
M0 V        &     $+8.87$     &       $+1.19$        &     $+1.41$        &    $+0.89$         &     $+1.76$        &    $+2.86$          &     $+3.50$        &    $+3.74$    &          0.59        \\
M1 V        &     $+9.56$     &       $+1.18$        &     $+1.47$        &    $+0.96$         &     $+1.98$        &    $+3.20$          &     $+3.77$        &    $+4.03$    &          0.54        \\
M2 V        &     $+10.17$     &       $+1.17$        &     $+1.50$        &    $+1.00$         &     $+2.14$        &    $+3.36$          &     $+3.94$        &    $+4.18$    &          0.45        \\
M3 V        &     $+11.01$     &       $+1.17$        &     $+1.55$        &    $+1.08$         &     $+2.45$        &    $+3.80$          &     $+4.38$        &    $+4.62$    &           0.33        \\
M4 V        &     $+12.80$     &       $+1.18$        &     $+1.67$        &    $+1.19$         &     $+2.75$        &    $+4.41$          &     $+4.96$        &    $+5.23$    &          0.24        \\
M5 V        &     $+14.20$     &       $+1.3:$        &     $+1.82$        &    $+1.41$         &     $+3.30$        &    $+5.13$          &     $+5.73$        &    $+6.00$    &           0.15        \\
M6 V        &     $+16.59$     &       $+1.3:$        &     $+2.03$        &    $+1.81$         &     $+3.93$        &    $+6.25$          &     $+6.86$        &    $+7.19$    &           0.11        \\
M7 V        &     $+17.84$     &       ...            &     $+2.15$        &    $+2.13$         &     $+4.51$        &    $+7.03$          &     $+7.64$        &    $+8.02$    &         0.10        \\
M8 V        &     $+18.72$     &       ...            &     $+2.15$        &    $+2.24$         &     $+4.57$        &    $+7.55$          &     $+8.23$        &    $+8.67$    &          0.09        \\
M9 V        &     $+19.39$     &       ...            &     $+2.15$        &    $+2.37$         &     $+4.61$        &    $+7.72$          &     $+8.45$        &    $+8.94$    &         0.08        \\
L0 V        &     $+19.65$     &       ...            &     ...            &    $+2.44$         &     $+4.66$        &    $+8.10$          &     $+8.82$        &    $+9.36$    &         0.08        \\
\hline
\end{tabular}
\caption{Spectral types, empirical $V$-band absolute magnitudes, intrinsic colour indices and masses of main-sequence stars  \label{tab:mainsequencecolours}}
}
\end{table*}

\end{document}